\PassOptionsToPackage{table}{xcolor}
\documentclass{article}

\PassOptionsToPackage{numbers, compress}{natbib}
\usepackage[preprint]{neurips_2026}

\usepackage[utf8]{inputenc} % allow utf-8 input
\usepackage[T1]{fontenc}    % use 8-bit T1 fonts
\usepackage{hyperref}       % hyperlinks
\usepackage{url}            % simple URL typesetting
\usepackage{booktabs}       % professional-quality tables
\usepackage{amsmath}        % \text, aligned environments, etc.
\usepackage{amsfonts}       % blackboard math symbols
\usepackage{nicefrac}       % compact symbols for 1/2, etc.
\usepackage{microtype}      % microtypography
\usepackage{xcolor}         % colors
\usepackage{colortbl}       % \cellcolor for heatmap-style tables
\usepackage{float}          % [H] placement for the appendix table stack
\usepackage{graphicx}       % include images

\usepackage{tikz}           % architecture diagram
\usetikzlibrary{positioning,arrows.meta,calc,fit}

\newcommand{\tablefont}{\fontsize{10pt}{12pt}\selectfont}
\newcommand{\chdie}{\raisebox{-0.45ex}{\includegraphics[height=2.6ex]{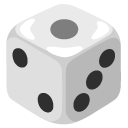}}}
\newcommand{\supp}{the Supplemental Materials (Appendices~\ref{app:losses}--\ref{app:runtime})}

\title{Helping Music Co-Creation Agents `Listen' Well: \\ Hierarchical Self-Supervised World Models for Understanding and Generation}

\author{%
  Scott H.~Hawley \\
  Department of Chemistry \& Physics\\
  Belmont University\\
  Nashville, TN, USA \\
  \texttt{scott.hawley@belmont.edu} \\
}

\begin{document}

\maketitle

\begin{abstract}
Collaborative music agents need internal representations rich enough to support both understanding and generation, yet flexible enough for a workflow where the human retains agency. We present a hierarchical self-supervised ``world model'' for symbolic music: a 2.55M-parameter Swin V2 encoder trained on MIDI piano-roll images with JEPA-style objectives (pitch- and time-shift equivariance, masked embedding prediction, and a distributional regularizer), using no labels and no music-theory vocabulary. Probing the frozen embeddings shows that the level at which a musical property becomes decodable tracks its musical time scale: phrase boundaries are read off the coarsest levels, note density and harmonic detail off the finest. Temporal and phrase structure emerge from the self-supervised objectives alone, while harmonic content must be asked for; a small chord-supervision head raises joint chord recovery from .18 to .54, and key detection, which is never supervised, from .16 to .70. Following the Representation AutoEncoder paradigm, a conditional flow-matching model stands in for a trained decoder, flowing in pixel space from PCA-reduced conditioning: it reproduces a target window at pixel F1 $0.996$, and the same per-level conditioning dropout that controls how far variations stray also enables graphical prompting for masked inpainting with no inpainting-specific sampler. The pipeline runs on CPU producing a suggestion in $2.8$ s, or $0.6$ s on Apple MPS, which we demonstrate in a live interactive demo. In concert with an LLM-based brain, these capabilities supply the core of a collaborative music creation agent in service of, rather than in place of, human agency.
\end{abstract}

\section{Paradigm: ``An AI Rick Rubin''}

Preserving human agency in the ``age of AI'' is a widespread concern particularly in the creative arts. 
This paper describes a system built to serve that goal, using small models and human domain expertise
to offer feedback and suggestions in a collaborative songwriting and production loop.  
The metaphor of an ``AI Rick Rubin'' captures the approach, after the producer who described his method in a 2023 interview with Anderson Cooper:
\vspace{-.2cm}

\begin{quote}
\textbf{Cooper:} Do you play instruments?\\
\textbf{Rubin:} Barely\ldots\ I have no technical ability. And I know nothing
about music.\\
\textbf{Cooper:} You must know something.\\
\textbf{Rubin:} Well, I know what I like and what I don't like. And I'm decisive
about what I like and what I don't like\ldots\ The confidence that I have in my
taste and my ability to express what I feel has proven helpful for
artists.~\cite{rubin2023sixtyminutes}
\end{quote}
\vspace{-.2cm}

\begin{figure}[tb]
\centering
\includegraphics[width=0.85\columnwidth]{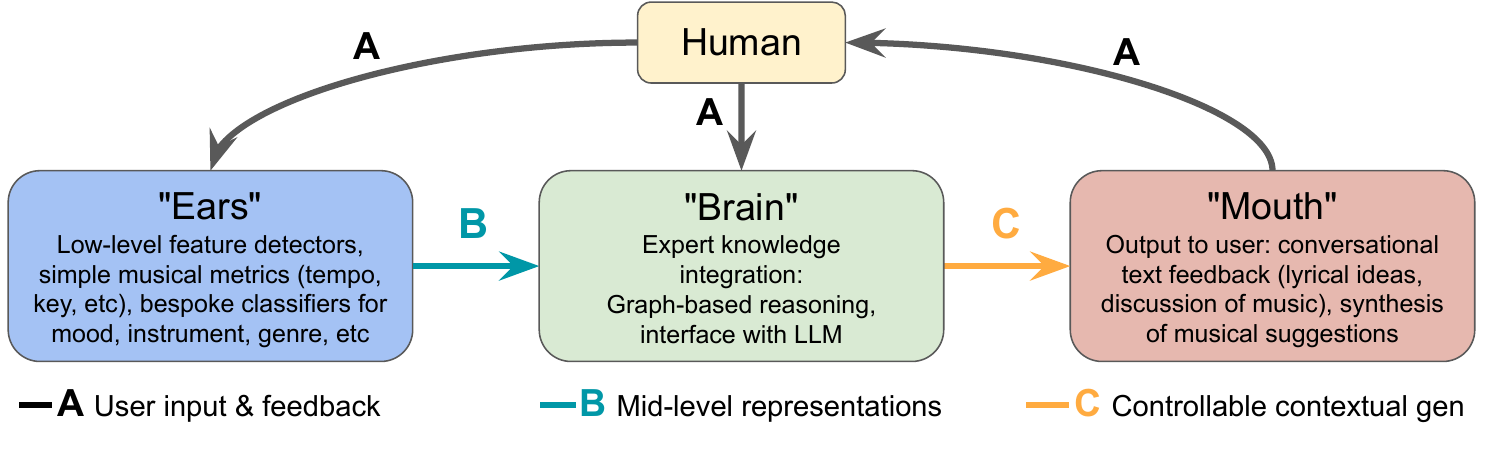}
\vspace{-0.4cm}
\caption{
The Ears/Brain/Mouth workflow for a conversational music co-writing agent. This paper
develops the Ears and Mouth: perceptual representations (Section~\ref{sec:ears}) rich
enough to drive controllable generation of music suggestions (Section~\ref{sec:mouth})
without task-specific fine-tuning. The Brain, an expert-knowledge reasoning layer, is
beyond the scope of this paper.
}
\label{fig:ears-brain-mouth}
\end{figure}

Our system listens carefully to the artist's musical ideas, processes them according to its own internal representations, 
and responds with verbal feedback and limited (generative) musical suggestions. The AI  never ``does it for'' the human: suggestions are left for the human to implement. In short, it's a good verbal articulator, but it ``barely'' plays any instruments. The structure of the system is illustrated in Figure~\ref{fig:ears-brain-mouth}; this paper describes essential components of the ``ears'' and ``mouth.''  That division of labor is deliberate. Large audio-language models vastly underperform on many simple music tasks
compared to smaller, targeted models~\cite{ma2025cmibench}, a finding that
persists in the symbolic domain~\cite{zhou2025abceval,lee2026howfar}. Systems 
that close the gap do so by attaching domain-specific perception
modules to the LLM~\cite{tan2026midillama}, the same role the ``ears''
play in our system.

Generative music models implicitly encode musical structure and style~\cite{castellon2021jukebox,evans2024stableaudio,min2023polyffusion}, but implicit encoding is not articulation: a co-writing partner must say what structures are present and what changes might improve a composition or arrangement. Machine agents could exchange feedback as mathematical signals, e.g. as gradients toward some endpoint, as in classifier guidance~\cite{dhariwal2021diffusion}, but a human collaborator cannot act on a gradient. 
Description and reproduction make different demands: representation learning and reconstruction are separate tasks, long regarded as a tradeoff~\cite{dieleman2025latents}.
Representation autoencoders (RAEs)~\cite{zheng2025rae} show the tradeoff is an
artifact of compression: their latents rearrange the input at nearly its original
dimensionality, with reconstruction handled separately rather than by the
representation objective. The pairing supports discriminative tasks and
high-quality generation alike. We adopt this paradigm for music, with a
conditional flow model standing in for the trained decoder: a single encoder,
trained with no reconstruction objective and then frozen, serving as the ears and
conditioning the mouth.

The internal representations are formed by a joint-embedding predictive
architecture (JEPA)~\cite{lecun2022path,assran2023ijepa}. JEPA has been applied
to music before: Stem-JEPA~\cite{riou2024stemjepa} predicted musical stem
compatibility from audio; Hachana and Rasheed~\cite{hachana2025using} adapted
JEPA to tokenized symbolic music with musically motivated masking of
instruments, pitch classes and octaves, and found the learned representation
dominated by positional information, which limited transfer to content tasks
such as genre and style; and, concurrently with this work,
Music-JEPA~\cite{wang2026musicjepa} learned an action-conditioned audio world
model from paired piano audio and performance data, while
ARIMA~\cite{anon2026arima} learned windowed latent-predictive representations of
symbolic music. Unlike these, our approach operates purely on symbolic (MIDI)
piano-roll images, is trained with a hierarchical, multi-level Swin V2 encoder
rather than a single-scale backbone, prevents representational collapse via
a LeJEPA-style distributional regularizer (SIGReg) rather than an EMA teacher, and
makes shift structure explicit in the objective rather than leaving it to be
absorbed by positional encodings.
This paper extends our earlier report~\cite{hawley2026midiraejepa}, which
details the architecture and training objectives; here we focus on the musical
and co-creative capabilities, and report improvements to the model and the generative pipeline together with an assessment of gains from supervised signals and additional datasets.

To maximize accessibility for GPU-poor musicians (who may also prefer not to send works-in-progress to a cloud service) 
we adopted a fundamental engineering constraint: \textit{timely inference on CPU-only runtimes}, demonstrated by a live demo linked from the upplemental materials.\footnote{Supplemental Materials: \href{https://drscotthawley.github.io/midi-rae-jepa-son}{\nolinkurl{drscotthawley.github.io/midi-rae-jepa-son}}}

\section{How: A Symbolic Music ``World Model''}

\begin{quote}\small\itshape
``It wasn't made by people who went to the music conservatory.
It was made by kids who felt something.''\hfill---Rick Rubin, on early hip-hop~\cite{rubin2023sixtyminutes}
\end{quote}
\vspace{-0.5\baselineskip}

\begin{figure}[tb]
\centering
% The PDF's page box is 720x405pt but its ink is only y=71.5..210.2 (Ghostscript
% -sDEVICE=bbox), so two thirds of the reserved height is blank. Trimmed to the ink;
% the previous \vspace{-1.3cm} was compensating for the lower half of that blank band.
% Regenerate from the SVG with:
%   rsvg-convert -f pdf -w 960 -h 540 -o pr-panorama-crops.pdf pr-panorama-crops.svg
% (-w/-h are px at 96dpi, so 960x540 px yields the 720x405 pt page these trim values
% assume; getting that wrong silently rescales the page and invalidates them).
\includegraphics[width=.8\columnwidth, trim=0 71.5bp 0 194.8bp, clip]{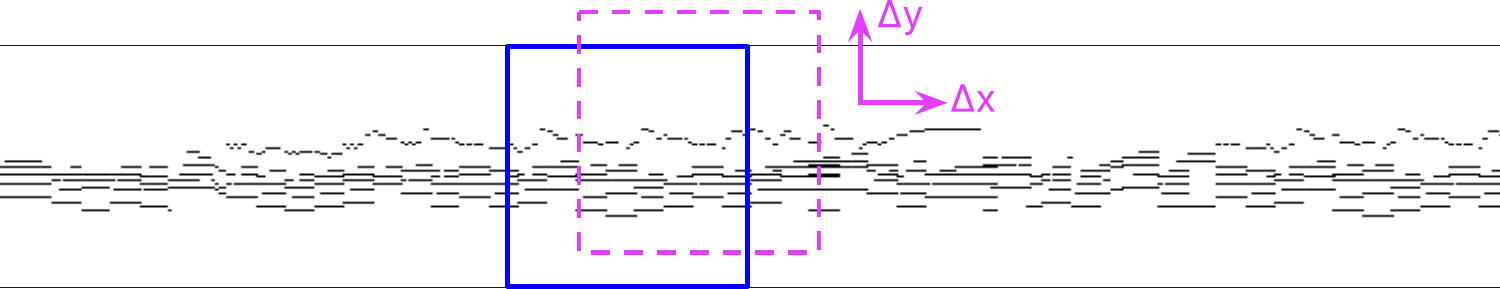}
\vspace{-.1cm}
\caption{Just as a vision world model pans and tilts a camera to scan a panoramic
scene, we take square crops of MIDI piano-roll images and compare crops
translated in time by $\Delta x$ time steps or in pitch by $\Delta y$
semitones. Given one crop and a shift vector $(\Delta x, \Delta y)$,
the model predicts the embedding of the shifted crop. Regions outside the image
are filled with blanks. Our design requirement of fast CPU execution fixes the
crop at 128$\times$128 pixels. Pixels are binary, so note velocity is discarded,
which is acceptable for the songwriting use case. 
Time is quantized to 32nd notes on a tempo-normalized grid, the finest grid at which a blank frame can separate two repeated notes in a binary image, so a crop spans four bars in 4/4, or eight seconds at 120 BPM.
}
\label{fig:panorama-crops}
\end{figure}

Rather than relying on supervised labels and the vocabulary of music theory, we train a self-supervised model that builds its own (hierarchical) representations of music, its own `feel' for musical structure. It does this by exploring the ``space'' of musical compositions, cf. Figure~\ref{fig:panorama-crops}. Human musical representations are hierarchical: notes, chords, phrases, song structure. Yet labeled data for that hierarchy is limited. Chord detectors and song-structure predictors can supply some of it, but we want to learn the hierarchy without labels, the way World Models in computer vision~\cite{ha2018worldmodels,garrido2024iwm} build semantic representations by predicting embeddings of one view of a scene from another. We do this on piano-roll images, partly to take advantage of the Swin V2 transformer~\cite{liu2021swin,liu2022swinv2}, a hierarchical representation learner already proven on images, and partly because a piano roll is a semantic proxy for an audio spectrogram, with perceptually aligned axes. 
Latents trained for reconstruction accuracy rarely become semantically
meaningful; they tend instead to resemble spatially downsampled copies of the
input. Following the RAE paradigm, ours is a latent model with \emph{no
compression}: the latents hold nearly as many numbers as the input (16,384
pixels $\rightarrow$ 16,128 floats), rearranged into a meaningful geometry
rather than a smaller one.

\begin{figure}[tb]
  \centering
  \resizebox{.88\columnwidth}{!}{\input{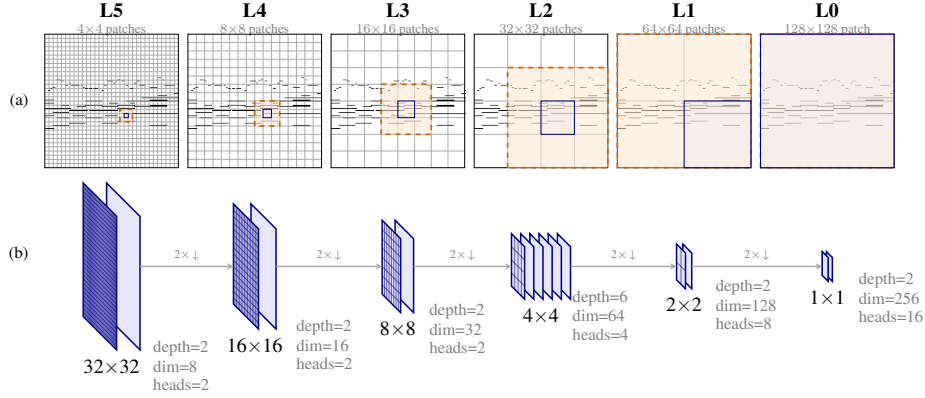}}
  \vspace{-.2cm}
  \caption{Our 2.55 million-parameter hierarchical Swin V2 Encoder. Model architecture hyperparameters were selected via ablation studies detailed in \supp.}
  \label{fig:arch}
\end{figure}

\textbf{Training objective.} The encoder is trained with a combined self-supervised objective,
$$\mathcal{L} = \lambda \mathcal{L}_{\text{equiv}} + (1-\lambda)\mathcal{L}_{\text{SIGReg}} + \lambda_{\text{MEP}}\mathcal{L}_{\text{MEP}} + \lambda_{\text{fact}}\mathcal{L}_{\text{fact}},$$
applied level-wise. $\mathcal{L}_{\text{equiv}}$ is an equivariance loss that pulls together or pushes apart embeddings of shifted crop pairs in proportion to the shift magnitude, so the latent space respects pitch/time translation structure rather than collapsing to shift-invariance.
$\mathcal{L}_{\text{SIGReg}}$ is LeJEPA's sketched isotropic Gaussian regularization~\cite{balestriero2025lejepa}, which applies an Epps--Pulley test along random projections to enforce an isotropic Gaussian prior, preventing representational collapse.
(We also tried VISReg~\cite{wu2026visreg}, which replaces the Epps--Pulley test with a sliced-Wasserstein distance, and saw no consistent improvement in our metrics.)
The equivariance and SIGReg losses are applied only at L0--L3: at L4 and L5,
SIGReg washes out the fine note-level structure those levels carry, and
patch-level equivariance is less meaningful at those scales.
$\mathcal{L}_{\text{MEP}}$ is an intra- and inter-level masked-embedding-prediction loss in the style of I-JEPA~\cite{assran2023ijepa}: an auxiliary predictor infers an EMA teacher's embeddings from the student's full context.
$\mathcal{L}_{\text{fact}}$ is a soft factorization loss which pulls pitch- and time-shift difference vectors toward parallel, anti-parallel, or orthogonal geometry depending on augmentation type.
It is applied only at L0--L2, enforcing equivariance out to larger shifts than $\mathcal{L}_{\text{equiv}}$ achieves alone~\cite{hawley2026midiraejepa}.
Full derivations and hyperparameters are given in \supp.

\section{Ears: What Does It Actually ``Perceive''?}
\label{sec:ears}

The ``ears'' of a full system such as that diagrammed in
Figure~\ref{fig:ears-brain-mouth} can partially consist of libraries of lightweight
audio feature detectors, predicting categories such as genre, instrument type,
and mood~\cite{essentia}, or metrics such as pitch and tempo~\cite{librosa}.
In addition, independent detectors for chords, keys, phrase boundaries 
and song structure are available.
For songwriting, we are concerned less with nuances of performance than with the notes the player intended. Discarding audio detail in favor of a structured MIDI representation therefore costs us little. 
Audio-to-MIDI
conversion~\cite{basic_pitch} recovers that intent in a compact, semantic form
already familiar from many music creators' workflows, and recent
improvements~\cite{mirelo_kyutai} have made it reliable enough to build on. 
We treat it as upstream of this work.
Our encoder is not meant to replace those small detectors. It learns representations that resist easy labeling, its own sense of ``feel,'' obtained only by exploring the space of musical compositions. To gauge what those representations correspond to musically, we turn to standard probes. Each probe below trains a small linear (or ridge) model on frozen, per-level embeddings, or measures a geometric property of the embedding space directly; none fine-tune the encoder.

\begin{table}[b]
\centering\tablefont
\caption{Probing frozen embeddings for musical structure. Each cell reports a
model's best value across its hierarchy levels (minimum for $\downarrow$
metrics), with the winning level in small type and a bar whose height is
proportional to that level's receptive-field size (L0 tallest); 
\scalebox{0.8}{\chdie{}} = chance level for that metric; 
bold = best model per row; cells are shaded per row with the
plasma colormap (yellow = best).
\textbf{Cross-song}: whether same-song embeddings sit closer together than
cross-song ones. \textbf{Time $R^2$}: how well the distance between two
embeddings predicts the temporal offset between their crops.
\textbf{Joint Chord}: root and chord quality jointly. \textbf{Chroma $R^2$}:
recovery of the local 12-bin pitch-class distribution.
\textbf{Phrase AP/AUC}: linear probe detecting whether a human-annotated
phrase boundary falls within half a bar of a crop's center, scored as average
precision and ROC-AUC.}
\label{tab:probes}
\begin{tabular}{lcccccccc}
\toprule
Metric & \chdie{} & DINOv2 & MRJ-48$\wedge$ & MRJS & +chords & +phrases & Lakh-1$\times$ & Lakh-4$\times$ \\
\midrule
Note Density $R^2$ $\uparrow$ & n/a & \cellcolor[RGB]{249,217,36}\textcolor{black}{.93\,{\scriptsize L1}\,\rule[-0.1ex]{1.4pt}{1.56ex}} & \cellcolor[RGB]{250,213,36}\textcolor{black}{.93\,{\scriptsize L5}\,\rule[-0.1ex]{1.4pt}{0.60ex}} & \cellcolor[RGB]{12,7,134}\textcolor{white}{.88\,{\scriptsize L5}\,\rule[-0.1ex]{1.4pt}{0.60ex}} & \cellcolor[RGB]{210,80,112}\textcolor{black}{.91\,{\scriptsize L5}\,\rule[-0.1ex]{1.4pt}{0.60ex}} & \cellcolor[RGB]{66,3,157}\textcolor{white}{.89\,{\scriptsize L5}\,\rule[-0.1ex]{1.4pt}{0.60ex}} & \cellcolor[RGB]{246,229,37}\textcolor{black}{.94\,{\scriptsize L5}\,\rule[-0.1ex]{1.4pt}{0.60ex}} & \cellcolor[RGB]{239,248,33}\textcolor{black}{\textbf{.94}\,{\scriptsize L5}\,\rule[-0.1ex]{1.4pt}{0.60ex}} \\
Cross-song $\downarrow$ & n/a & \cellcolor[RGB]{208,77,115}\textcolor{black}{.68\,{\scriptsize L1}\,\rule[-0.1ex]{1.4pt}{1.56ex}} & \cellcolor[RGB]{194,61,128}\textcolor{white}{.69\,{\scriptsize L5}\,\rule[-0.1ex]{1.4pt}{0.60ex}} & \cellcolor[RGB]{12,7,134}\textcolor{white}{.80\,{\scriptsize L3}\,\rule[-0.1ex]{1.4pt}{1.08ex}} & \cellcolor[RGB]{143,13,163}\textcolor{white}{.73\,{\scriptsize L1}\,\rule[-0.1ex]{1.4pt}{1.56ex}} & \cellcolor[RGB]{239,248,33}\textcolor{black}{\textbf{.57}\,{\scriptsize L3}\,\rule[-0.1ex]{1.4pt}{1.08ex}} & \cellcolor[RGB]{250,157,58}\textcolor{black}{.62\,{\scriptsize L4}\,\rule[-0.1ex]{1.4pt}{0.84ex}} & \cellcolor[RGB]{76,2,161}\textcolor{white}{.77\,{\scriptsize L5}\,\rule[-0.1ex]{1.4pt}{0.60ex}} \\
Time $R^2$ $\uparrow$ & n/a & \cellcolor[RGB]{84,1,163}\textcolor{white}{.23\,{\scriptsize L1}\,\rule[-0.1ex]{1.4pt}{1.56ex}} & \cellcolor[RGB]{194,61,128}\textcolor{white}{.24\,{\scriptsize L2}\,\rule[-0.1ex]{1.4pt}{1.32ex}} & \cellcolor[RGB]{239,248,33}\textcolor{black}{\textbf{.26}\,{\scriptsize L4}\,\rule[-0.1ex]{1.4pt}{0.84ex}} & \cellcolor[RGB]{212,82,110}\textcolor{black}{.24\,{\scriptsize L1}\,\rule[-0.1ex]{1.4pt}{1.56ex}} & \cellcolor[RGB]{210,80,112}\textcolor{black}{.24\,{\scriptsize L2}\,\rule[-0.1ex]{1.4pt}{1.32ex}} & \cellcolor[RGB]{243,134,73}\textcolor{black}{.25\,{\scriptsize L2}\,\rule[-0.1ex]{1.4pt}{1.32ex}} & \cellcolor[RGB]{12,7,134}\textcolor{white}{.22\,{\scriptsize L3}\,\rule[-0.1ex]{1.4pt}{1.08ex}} \\
Root Note $\uparrow$ & \cellcolor[RGB]{12,7,134}\textcolor{white}{.08} & \cellcolor[RGB]{143,13,163}\textcolor{white}{.24\,{\scriptsize L0}\,\rule[-0.1ex]{1.4pt}{1.80ex}} & \cellcolor[RGB]{154,21,158}\textcolor{white}{.25\,{\scriptsize L5}\,\rule[-0.1ex]{1.4pt}{0.60ex}} & \cellcolor[RGB]{149,17,161}\textcolor{white}{.24\,{\scriptsize L5}\,\rule[-0.1ex]{1.4pt}{0.60ex}} & \cellcolor[RGB]{239,248,33}\textcolor{black}{\textbf{.59}\,{\scriptsize L2}\,\rule[-0.1ex]{1.4pt}{1.32ex}} & \cellcolor[RGB]{133,6,166}\textcolor{white}{.22\,{\scriptsize L5}\,\rule[-0.1ex]{1.4pt}{0.60ex}} & \cellcolor[RGB]{132,5,166}\textcolor{white}{.22\,{\scriptsize L5}\,\rule[-0.1ex]{1.4pt}{0.60ex}} & \cellcolor[RGB]{162,28,154}\textcolor{white}{.26\,{\scriptsize L5}\,\rule[-0.1ex]{1.4pt}{0.60ex}} \\
Joint Chord $\uparrow$ & \cellcolor[RGB]{12,7,134}\textcolor{white}{.05} & \cellcolor[RGB]{126,3,167}\textcolor{white}{.17\,{\scriptsize L1}\,\rule[-0.1ex]{1.4pt}{1.56ex}} & \cellcolor[RGB]{124,2,167}\textcolor{white}{.17\,{\scriptsize L5}\,\rule[-0.1ex]{1.4pt}{0.60ex}} & \cellcolor[RGB]{132,5,166}\textcolor{white}{.18\,{\scriptsize L5}\,\rule[-0.1ex]{1.4pt}{0.60ex}} & \cellcolor[RGB]{239,248,33}\textcolor{black}{\textbf{.54}\,{\scriptsize L3}\,\rule[-0.1ex]{1.4pt}{1.08ex}} & \cellcolor[RGB]{126,3,167}\textcolor{white}{.18\,{\scriptsize L5}\,\rule[-0.1ex]{1.4pt}{0.60ex}} & \cellcolor[RGB]{129,4,167}\textcolor{white}{.18\,{\scriptsize L5}\,\rule[-0.1ex]{1.4pt}{0.60ex}} & \cellcolor[RGB]{143,13,163}\textcolor{white}{.20\,{\scriptsize L5}\,\rule[-0.1ex]{1.4pt}{0.60ex}} \\
Key Detection $\uparrow$ & \cellcolor[RGB]{12,7,134}\textcolor{white}{.05} & \cellcolor[RGB]{109,0,168}\textcolor{white}{.19\,{\scriptsize L0}\,\rule[-0.1ex]{1.4pt}{1.80ex}} & \cellcolor[RGB]{106,0,167}\textcolor{white}{.18\,{\scriptsize L5}\,\rule[-0.1ex]{1.4pt}{0.60ex}} & \cellcolor[RGB]{92,0,165}\textcolor{white}{.16\,{\scriptsize L5}\,\rule[-0.1ex]{1.4pt}{0.60ex}} & \cellcolor[RGB]{239,248,33}\textcolor{black}{\textbf{.70}\,{\scriptsize L3}\,\rule[-0.1ex]{1.4pt}{1.08ex}} & \cellcolor[RGB]{100,0,167}\textcolor{white}{.17\,{\scriptsize L5}\,\rule[-0.1ex]{1.4pt}{0.60ex}} & \cellcolor[RGB]{87,1,164}\textcolor{white}{.15\,{\scriptsize L5}\,\rule[-0.1ex]{1.4pt}{0.60ex}} & \cellcolor[RGB]{104,0,167}\textcolor{white}{.18\,{\scriptsize L5}\,\rule[-0.1ex]{1.4pt}{0.60ex}} \\
Chroma $R^2$ $\uparrow$ & \cellcolor[RGB]{12,7,134}\textcolor{white}{.00} & \cellcolor[RGB]{186,52,135}\textcolor{white}{.36\,{\scriptsize L1}\,\rule[-0.1ex]{1.4pt}{1.56ex}} & \cellcolor[RGB]{234,115,86}\textcolor{black}{.54\,{\scriptsize L4}\,\rule[-0.1ex]{1.4pt}{0.84ex}} & \cellcolor[RGB]{250,159,58}\textcolor{black}{.65\,{\scriptsize L5}\,\rule[-0.1ex]{1.4pt}{0.60ex}} & \cellcolor[RGB]{239,248,33}\textcolor{black}{\textbf{.83}\,{\scriptsize L4}\,\rule[-0.1ex]{1.4pt}{0.84ex}} & \cellcolor[RGB]{251,209,36}\textcolor{black}{.75\,{\scriptsize L5}\,\rule[-0.1ex]{1.4pt}{0.60ex}} & \cellcolor[RGB]{250,213,36}\textcolor{black}{.76\,{\scriptsize L5}\,\rule[-0.1ex]{1.4pt}{0.60ex}} & \cellcolor[RGB]{246,145,66}\textcolor{black}{.61\,{\scriptsize L5}\,\rule[-0.1ex]{1.4pt}{0.60ex}} \\
Phrase AP $\uparrow$ & \cellcolor[RGB]{12,7,134}\textcolor{white}{.17} & \cellcolor[RGB]{147,16,161}\textcolor{white}{.21\,{\scriptsize L0}\,\rule[-0.1ex]{1.4pt}{1.80ex}} & \cellcolor[RGB]{253,188,42}\textcolor{black}{.28\,{\scriptsize L1}\,\rule[-0.1ex]{1.4pt}{1.56ex}} & \cellcolor[RGB]{253,175,49}\textcolor{black}{.27\,{\scriptsize L0}\,\rule[-0.1ex]{1.4pt}{1.80ex}} & \cellcolor[RGB]{229,107,92}\textcolor{black}{.25\,{\scriptsize L2}\,\rule[-0.1ex]{1.4pt}{1.32ex}} & \cellcolor[RGB]{239,248,33}\textcolor{black}{\textbf{.29}\,{\scriptsize L2}\,\rule[-0.1ex]{1.4pt}{1.32ex}} & \cellcolor[RGB]{251,209,36}\textcolor{black}{.28\,{\scriptsize L1}\,\rule[-0.1ex]{1.4pt}{1.56ex}} & \cellcolor[RGB]{246,142,68}\textcolor{black}{.26\,{\scriptsize L0}\,\rule[-0.1ex]{1.4pt}{1.80ex}} \\
Phrase AUC $\uparrow$ & \cellcolor[RGB]{12,7,134}\textcolor{white}{.50} & \cellcolor[RGB]{227,104,94}\textcolor{black}{.57\,{\scriptsize L0}\,\rule[-0.1ex]{1.4pt}{1.80ex}} & \cellcolor[RGB]{244,234,38}\textcolor{black}{.61\,{\scriptsize L2}\,\rule[-0.1ex]{1.4pt}{1.32ex}} & \cellcolor[RGB]{244,234,38}\textcolor{black}{.61\,{\scriptsize L0}\,\rule[-0.1ex]{1.4pt}{1.80ex}} & \cellcolor[RGB]{240,128,77}\textcolor{black}{.58\,{\scriptsize L0}\,\rule[-0.1ex]{1.4pt}{1.80ex}} & \cellcolor[RGB]{239,248,33}\textcolor{black}{\textbf{.62}\,{\scriptsize L0}\,\rule[-0.1ex]{1.4pt}{1.80ex}} & \cellcolor[RGB]{250,216,36}\textcolor{black}{.61\,{\scriptsize L1}\,\rule[-0.1ex]{1.4pt}{1.56ex}} & \cellcolor[RGB]{242,240,38}\textcolor{black}{.61\,{\scriptsize L0}\,\rule[-0.1ex]{1.4pt}{1.80ex}} \\
\bottomrule
\end{tabular}
\end{table}

We call the model introduced here MIDI-RAE-JEPA-SON (MRJS), a successor of our earlier report's MIDI-RAE-JEPA~\cite{hawley2026midiraejepa}, which appears in Table~\ref{tab:probes} as MRJ-48$\wedge$ ($\Delta t_{\max}{=}48$, factorization loss active), alongside a DINOv2~\cite{dinov2} baseline and a chance column (\chdie{}). Unlike MRJ-48$\wedge$, whose pipeline components were trained against differing train/test splits on a dataset since found to contain corruption, MRJS and every downstream component share a single unified split on corrected data. That column is therefore historical reference, not a controlled comparison.
The table reports each probe at the hierarchy level where it scores best, and those levels carry as much information as the scores. Phrase boundaries peak at the coarse end (L0--L2) while note density and harmonic content peak at the fine end (L4--L5), so the level at which a property becomes decodable tracks the musical time scale on which it operates. Temporal offset is the weakest probe throughout, .22--.26 for every model, which is expected: a piano roll is nearly stationary along the time axis, so sliding a crop sideways yields more of the same, whereas pitch is absolute and a vertical shift moves material into a different register. 
We read the harmony scores as capacity probes, not as downstream performance; small dedicated chord detectors do that job far better.  
The DINOv2 baseline is instructive: on properties that reduce to image
statistics it is competitive or better, reaching .93 on note density and
separating songs more sharply than MRJS, but it falls away on the musically
specific probes, recovering chroma at .36 against .65 and phrase boundaries at
.21 AP against .27. 
%A generic vision backbone reads the picture; the musical structure has to be trained for.

To assess the effects of dataset variation and size, we trained encoders on subsets of the Lakh MIDI Dataset~\cite{raffel2016lakh} containing $1\times$ and $4\times$ as many songs as POP909. Both transfer well: on the POP909 probes they match or exceed MRJS on note density and cross-song separation. At matched training budget, however, the $4\times$ subset gave no consistent gain over $1\times$, and degraded chroma (.76 to .61) and cross-song separation (.62 to .77).
We also explored adding auxiliary supervised objectives to the self-supervised recipe: chord recognition, using POP909's chord annotations, and phrase-boundary prediction, using the human-verified annotations of Dai et al.~\cite{dai2020structure}, each applied as a small linear head on the coarse levels during training. Chord supervision makes harmonic content linearly decodable where it was barely present: root identification rises from .24 to .59, joint chord from .18 to .54, and key detection, which was never supervised, from .16 to .70. It also moves the seat 
of that information from L5 (near note-level) to the coarser and more ``abstract'' L2--L3. The cost is small: phrase AP falls from .27 to .25 and temporal offset from .26 to .24, while note density and cross-song separation both improve. Phrase supervision helps less, but is not without effect: boundary detection rises only slightly (AP .27 to .29, AUC .61 to .62), while cross-song separation improves markedly (.80 to .57). Harmonic content, then, has to be asked for; temporal and phrase structure largely emerge from the self-supervised objectives alone.
Once the inputs are encoded, they can be sent to the ``brain'' as well as used as conditioning signals for the mouth described next.

\section{Mouth: Making Suggestions}
\label{sec:mouth}

Verbal feedback and lyrical suggestions can be handled by a properly harnessed LLM.
Musical suggestions benefit from their own generative model conditioned on the 
idea the user already has, as in an
inpainting task where we replace a melody or an accompaniment, or suggest a different continuation.
Hachana and Rasheed~\cite{hachana2025using} anticipate that a reliable JEPA music representation could enable generative applications; this section realizes one.
Autoregressive transformers generate left to right, so filling a hole in the middle of an existing passage is not something they do natively, and must be trained for specifically~\cite{tan2022melodyinfilling}. Diffusion models over piano-roll images, by contrast, support inpainting at inference
time essentially for free~\cite{min2023polyffusion,hawley2024picturesofmidi}, and in a
form that admits \textit{graphical prompts}: the user draws on the piano roll where they want
new notes, and the model generates notes suiting the surrounding material.
But the price of this power is inference cost. Diffusion inpainting needs many
integration steps, and our own earlier system, Pictures of MIDI~\cite{hawley2024picturesofmidi}, was large and slow
enough to require a GPU. That is disqualifying under our design constraint.
We therefore moved to flow matching~\cite{lipman2023flow}, which can generate in very
few integration steps: the smooth flow admits higher-order integration schemes such as RK4, and optimal-transport pairing~\cite{tong2024improving} during training straightens the trajectories,
making even Euler steps sufficient.

Recent work has favored flowing in pixel space rather than in the latent space~\cite{chen2025pixelflow,lu2026pixelmeanflow}. We do the same: we flow from noise in pixel space, conditioned on embeddings of the user's existing musical idea.
We said earlier that there is no compression in the embeddings; however, 
we find it sufficient to reduce the conditioning signal via PCA, keeping at least 
90\% of the variance per level, resulting in a reduction of over $3\times$ (16,128 $\rightarrow$ 4,835 floats).
This still reconstructs the input closely (F1=$0.996$), 
yet the goal is variations rather than copies. How far those variations stray is
controllable: guidance strength sets how tightly a sample follows its conditioning, and
dropping conditioning at different levels loosens it further. That dropout also points
to how we handle inpainting.
To inpaint, we apply spatial dropout to the PCA conditioning at the locations the
user painted. 
%The flow was trained with exactly this dropout, so it fills the gap from surrounding context without needing an inpainting-specific sampler; in this
% respect it resembles the CRASH~\cite{rouard2021crashrawaudioscorebased} method for diffusion models, and we adopt none of the flow-specific schemes~\cite{pokle2024trainingfree, pnp_flow}. 
The flow was trained with exactly this dropout, so it fills the gap from surrounding context without needing an inpainting-specific sampler, unlike sampling-time schemes for diffusion models~\cite{rouard2021crashrawaudioscorebased}, and we adopt none of the flow-specific schemes~\cite{pokle2024trainingfree, pnp_flow}.
Varying the dropout probability per level sets how coarse or fine the replacement is: dropping only the fine levels rewrites notes while
keeping the harmonic frame, dropping every level rewrites the passage.

\begin{figure}[tb]
\centering
\includegraphics[width=.85\textwidth]{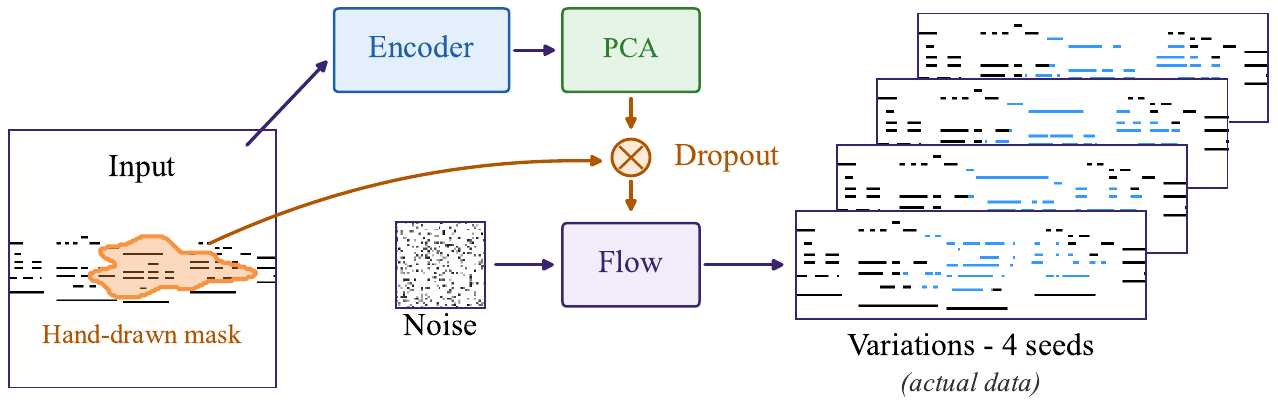}
\caption{
Unified generation and inpainting. The encoder's PCA-reduced conditioning maps pass through a
dropout stage that zeroes patches under a hand-drawn mask, at user-chosen
strengths per level; a flow-matching model then generates a window from noise,
guided only by the surviving conditioning (10 Euler steps, guidance 1). Outside
the mask the output is overwritten with the input, preserving untouched music
exactly. With no mask, the same pipeline performs ordinary conditioned
generation, varied by seed or guidance strength. The piano rolls are real data: one
POP909 excerpt and four generated variations (cropped to the occupied pitch range
for display).
}
\label{fig:inpaint-pipeline}
\end{figure}

Figure~\ref{fig:inpaint-pipeline} shows four inpainted variations of a real
excerpt, and Table~\ref{tab:suggestions} quantifies both generation and
inpainting. Infilled notes generally fit the surrounding music, 
though they follow it less closely than larger diffusion models~\cite{hawley2024picturesofmidi}, and some land off. 
That manifests as variability in how much material the model commits inside the mask.
The paradigm absorbs it: an AI Rick Rubin ``barely'' plays the instrument, and a
suggestion need only convey an idea the user implements themselves.

\begin{table}[tb]
\centering
\small
\caption{
\textbf{(a)} Conditioned on a real window's own embeddings, the flow reproduces
it nearly pixel-perfectly with no post-hoc alignment: the PCA-reduced
conditioning retains enough to rebuild the input. \textbf{(b)} Note density
restored in the masked region under the dropout pipeline of
Fig.~\ref{fig:inpaint-pipeline}: the full input is always encoded, and
conditioning patches under the mask are dropped at the stated per-level
strengths. Restoration falls monotonically with dropout strength; spreads are
$\pm 1$--$23\%$. Both panels: $10$ Euler steps, guidance $1.0$, over $6$ songs
$\times\,3$ seeds, with $3$ masks in (b).
}
\begin{tabular}{lc}
\toprule
\emph{(a) Reconstruction, pixel F1} & $\mathbf{0.996 \pm 0.004}$ \\
\midrule
\multicolumn{2}{l}{\emph{(b) Note density restored in masked region vs.\ dropout strength}} \\
\quad no dropout (reconstruction)     & $100\%$ \\
\quad fine levels (L4--L5) @ $0.85$   & $96\%$  \\
\quad fine levels (L4--L5) @ $1.0$    & $76\%$  \\
\quad all levels @ $1.0$              & $53\%$  \\
\bottomrule
\end{tabular}
\label{tab:suggestions}
\end{table}

\section{Outro: What To Do With This?}

We built a live demo\footnote{Best experienced firsthand:
\href{https://drscotthawley-midi-rae-jepa-son.hf.space}{\nolinkurl{drscotthawley-midi-rae-jepa-son.hf.space}}}
that serves not only as an interactive probe of the model's representational
capacity, but as a \textit{super fun} app that shows how this kind of control
feels in practice. Encoding a $128\times128$ window takes $8.6$ ms on two CPU
threads, faster than Apple's Metal backend ($19.9$ ms). A suggestion (10 Euler
steps at guidance $1.0$, one function evaluation per step) takes $3.8$ s on two
CPU threads and $2.8$ s on a full CPU (M1 Max, no MPS), within our ``timely''
design target; MPS drops this to $0.6$ s, a laptop RTX 4090 to $0.10$ s.
Sampling is over $99\%$ of these times, so latency scales with function
evaluations. Because the model is likelier to under-fill a mask than over-fill
it, the demo streams several variations ordered by how many new notes each
contains, so the next is ready while the user auditions the last.
Future work could extend the temporal field by treating the coarsest levels as
tokens in a sequence model, or by folding piano-roll images into large squares
as in~\cite{hawley2024picturesofmidi}.
Adding channels to encode note onsets, velocities, and multiple instruments would lead to richer musical capabilities, and an onset channel would free the separator frames that currently halve the window's span.
While this system is intended for interacting with humans, nothing prevents
someone from wiring it to a production system and closing the loop entirely.
However, we strongly believe in AI for enhancing humans' experience of
creativity, not for replacing it.

\begin{ack}
We thank Razer Corporation and TwinOS for providing the GPU laptops used for
some of the computations in this work.
\end{ack}

\bibliographystyle{plain}
\bibliography{midi_rae_refs}

\begin{thebibliography}{10}

\bibitem{assran2023ijepa}
Mahmoud Assran, Quentin Duval, Ishan Misra, Piotr Bojanowski, Pascal Vincent,
  Michael Rabbat, Yann LeCun, and Nicolas Ballas.
\newblock Self-supervised learning from images with a joint-embedding
  predictive architecture.
\newblock In {\em Conference on Computer Vision and Pattern Recognition}, 2023.

\bibitem{balestriero2025lejepa}
Randall Balestriero and Yann LeCun.
\newblock {LeJEPA}: Provable and scalable self-supervised learning without the
  heuristics, 2025.

\bibitem{basic_pitch}
Rachel~M. Bittner, Juan~Jos{\'e} Bosch, David Rubinstein, Gabriel
  Meseguer-Brocal, and Sebastian Ewert.
\newblock A lightweight instrument-agnostic model for polyphonic note
  transcription and multipitch estimation.
\newblock In {\em Proceedings of the IEEE International Conference on
  Acoustics, Speech, and Signal Processing (ICASSP)}, Singapore, 2022.

\bibitem{essentia}
Dmitry Bogdanov, Nicolas Wack, Emilia G{\'o}mez, Sankalp Gulati, Perfecto
  Herrera, Oscar Mayor, Gerard Roma, Justin Salamon, Jos{\'e}~R. Zapata, and
  Xavier Serra.
\newblock Essentia: An audio analysis library for music information retrieval.
\newblock In {\em Proceedings of the 14th International Society for Music
  Information Retrieval Conference (ISMIR)}, pages 493--498, Curitiba, Brazil,
  2013.

\bibitem{caron2021dino}
Mathilde Caron, Hugo Touvron, Ishan Misra, Herv{\'e} J{\'e}gou, Julien Mairal,
  Piotr Bojanowski, and Armand Joulin.
\newblock Emerging properties in self-supervised vision transformers.
\newblock In {\em International Conference on Computer Vision}, 2021.

\bibitem{castellon2021jukebox}
Rodrigo Castellon, Chris Donahue, and Percy Liang.
\newblock Codified audio language modeling learns useful representations for
  music information retrieval.
\newblock In {\em Proceedings of the International Society for Music
  Information Retrieval Conference (ISMIR)}, 2021.

\bibitem{chen2020musicsketchnet}
Ke~Chen, Gus Xia, and Shlomo Dubnov.
\newblock {Music SketchNet}: Controllable music generation via factorized
  representations of pitch and rhythm.
\newblock {\em arXiv preprint arXiv:2008.01291}, 2020.

\bibitem{chen2025pixelflow}
Shoufa Chen, Chongjian Ge, Shilong Zhang, Peize Sun, and Ping Luo.
\newblock Pixelflow: Pixel-space generative models with flow, 2025.

\bibitem{dai2020structure}
Shuqi Dai, Huan Zhang, and Roger~B. Dannenberg.
\newblock Automatic analysis and influence of hierarchical structure on melody,
  rhythm and harmony in popular music.
\newblock In {\em Proceedings of the Joint Conference on AI Music Creativity
  (AIMC)}, 2020.
\newblock arXiv:2010.07518. Human-verified phrase-level structure annotations
  for POP909 at
  \url{https://github.com/Dsqvival/hierarchical-structure-analysis}.

\bibitem{dhariwal2021diffusion}
Prafulla Dhariwal and Alex Nichol.
\newblock Diffusion models beat {GANs} on image synthesis.
\newblock In {\em Advances in Neural Information Processing Systems}, 2021.

\bibitem{dieleman2025latents}
Sander Dieleman.
\newblock Generative modelling in latent space, 2025.

\bibitem{evans2024stableaudio}
Zach Evans, CJ~Carr, Josiah Taylor, Scott~H. Hawley, and Jordi Pons.
\newblock Fast timing-conditioned latent audio diffusion.
\newblock In {\em Proceedings of the International Conference on Machine
  Learning (ICML)}, 2024.

\bibitem{garrido2024iwm}
Quentin Garrido, Mahmoud Assran, Nicolas Ballas, Adrien Bardes, Laurent Najman,
  and Yann LeCun.
\newblock Learning and leveraging world models in visual representation
  learning.
\newblock {\em arXiv preprint arXiv:2403.00504}, 2024.

\bibitem{ha2018worldmodels}
David Ha and J{\"u}rgen Schmidhuber.
\newblock World models.
\newblock {\em arXiv preprint arXiv:1803.10122}, 2018.

\bibitem{hachana2025using}
Rafik Hachana and Bader Rasheed.
\newblock Using a joint-embedding predictive architecture for symbolic music
  understanding.
\newblock In {\em NeurIPS 2025 Workshop on AI for Music}, 2025.

\bibitem{hawley2024picturesofmidi}
Scott~H. Hawley.
\newblock Pictures of midi: Controlled music generation via graphical prompts
  for image-based diffusion inpainting, 2024.

\bibitem{hawley2026midiraejepa}
Scott~H. Hawley.
\newblock {MIDI-RAE-JEPA}: {Hierarchical} representation learning and
  generation for symbolic music, 2026.

\bibitem{emopia}
Hsiao-Tzu Hung, Joann Ching, Seungheon Doh, Nabin Kim, Juhan Nam, and Yi-Hsuan
  Yang.
\newblock {EMOPIA}: A multi-modal pop piano dataset for emotion recognition and
  emotion-based music generation.
\newblock In {\em Proc. Int. Society for Music Information Retrieval Conf.},
  2021.

\bibitem{lee2026howfar}
Deepak Kumar, Emmanouil Karystinaios, Gerhard Widmer, and Markus Schedl.
\newblock How far can pretrained {LLMs} go in symbolic music? controlled
  comparisons of supervised and preference-based adaptation.
\newblock {\em arXiv preprint arXiv:2601.22764}, 2026.

\bibitem{lecun2022path}
Yann LeCun.
\newblock A path towards autonomous machine intelligence.
\newblock {\em OpenReview preprint}, 2022.
\newblock Version 0.9.2.

\bibitem{lipman2023flow}
Yaron Lipman, Ricky T.~Q. Chen, Heli Ben-Hamu, Maximilian Nickel, and Matthew
  Le.
\newblock Flow matching for generative modeling.
\newblock In {\em The Eleventh International Conference on Learning
  Representations}, 2023.

\bibitem{liu2022swinv2}
Ze~Liu, Han Hu, Yutong Lin, Zhuliang Yao, Zhenda Xie, Yixuan Wei, Jia Ning, Yue
  Cao, Zheng Zhang, Li~Dong, et~al.
\newblock Swin transformer {V2}: Scaling up capacity and resolution.
\newblock In {\em Conference on Computer Vision and Pattern Recognition}, 2022.

\bibitem{liu2021swin}
Ze~Liu, Yutong Lin, Yue Cao, Han Hu, Yixuan Wei, Zheng Zhang, Stephen Lin, and
  Baining Guo.
\newblock Swin transformer: Hierarchical vision transformer using shifted
  windows.
\newblock In {\em International Conference on Computer Vision}, 2021.

\bibitem{lu2026pixelmeanflow}
Yiyang Lu, Susie Lu, Qiao Sun, Hanhong Zhao, Zhicheng Jiang, Xianbang Wang,
  Tianhong Li, Zhengyang Geng, and Kaiming He.
\newblock One-step latent-free image generation with pixel mean flows, 2026.

\bibitem{ma2025cmibench}
Yinghao Ma, Siyou Li, Juntao Yu, Emmanouil Benetos, and Akira Maezawa.
\newblock {CMI-Bench}: A comprehensive benchmark for evaluating music
  instruction following.
\newblock In {\em Proceedings of the 26th International Society for Music
  Information Retrieval Conference (ISMIR)}, 2025.

\bibitem{pnp_flow}
S\'{e}gol\`{e}ne Martin, Anne Gagneux, Paul Hagemann, and Gabriele Steidl.
\newblock Pnp-flow: Plug-and-play image restoration with flow matching.
\newblock In Y.~Yue, A.~Garg, N.~Peng, F.~Sha, and R.~Yu, editors, {\em
  {I}nternational {C}onference on {R}epresentation {L}earning}, volume 2025,
  pages 45466--45492, 2025.

\bibitem{librosa}
Brian McFee, Colin Raffel, Dawen Liang, Daniel P.~W. Ellis, Matt McVicar, Eric
  Battenberg, and Oriol Nieto.
\newblock librosa: Audio and music signal analysis in {P}ython.
\newblock In {\em Proceedings of the 14th Python in Science Conference
  (SciPy)}, pages 18--24, 2015.

\bibitem{min2023polyffusion}
Lejun Min, Junyan Jiang, Gus Xia, and Jingwei Zhao.
\newblock Polyffusion: A diffusion model for polyphonic score generation with
  internal and external controls.
\newblock In {\em Proceedings of the 24th International Society for Music
  Information Retrieval Conference (ISMIR)}, 2023.

\bibitem{dinov2}
Maxime Oquab, Timoth{\'e}e Darcet, Th{\'e}o Moutakanni, Huy Vo, Marc
  Szafraniec, Vasil Khalidov, Pierre Fernandez, Daniel Haziza, Francisco Massa,
  Alaaeldin El-Nouby, et~al.
\newblock {DINOv2}: Learning robust visual features without supervision.
\newblock {\em arXiv preprint arXiv:2304.07193}, 2024.

\bibitem{pokle2024trainingfree}
Ashwini Pokle, Matthew~J. Muckley, Ricky T.~Q. Chen, and Brian Karrer.
\newblock Training-free linear image inverses via flows.
\newblock {\em {T}ransactions on {M}achine {L}earning {R}esearch}, 2024.

\bibitem{raffel2016lakh}
Colin Raffel.
\newblock {\em Learning-Based Methods for Comparing Sequences, with
  Applications to Audio-to-{MIDI} Alignment and Matching}.
\newblock PhD thesis, Columbia University, 2016.

\bibitem{riou2024stemjepa}
Alain Riou, Stefan Lattner, Ga{\"e}tan Hadjeres, Michael Anslow, and Geoffroy
  Peeters.
\newblock {Stem-JEPA}: A joint-embedding predictive architecture for musical
  stem compatibility estimation.
\newblock In {\em International Society for Music Information Retrieval
  Conference}, 2024.

\bibitem{rouard2021crashrawaudioscorebased}
Simon Rouard and Gaëtan Hadjeres.
\newblock Crash: Raw audio score-based generative modeling for controllable
  high-resolution drum sound synthesis, 2021.

\bibitem{mirelo_kyutai}
Simon Rouard, Michael Krause, Axel Roebel, Carl-Johann Simon-Gabriel, and
  Alexandre D{\'e}fossez.
\newblock Muscriptor: An open model for multi-instrument music transcription,
  2026.

\bibitem{rubin2023sixtyminutes}
Rick Rubin.
\newblock {Rick Rubin}: The 60 minutes interview.
\newblock \emph{60 Minutes}, CBS News. Interview by Anderson Cooper, January
  2023.

\bibitem{tan2022melodyinfilling}
Chih-Pin Tan, Alvin W.~Y. Su, and Yi-Hsuan Yang.
\newblock Melody infilling with user-provided structural context.
\newblock In {\em Proceedings of the 23rd International Society for Music
  Information Retrieval Conference (ISMIR)}, Bengaluru, India, 2022.

\bibitem{tong2024improving}
Alexander Tong, Kilian FATRAS, Nikolay Malkin, Guillaume Huguet, Yanlei Zhang,
  Jarrid Rector-Brooks, Guy Wolf, and Yoshua Bengio.
\newblock Improving and generalizing flow-based generative models with
  minibatch optimal transport.
\newblock {\em Transactions on Machine Learning Research}, 2024.
\newblock Expert Certification.

\bibitem{wang2026musicjepa}
Ziyu Wang, Kun Fang, and Yann LeCun.
\newblock {Music-JEPA}: Learning a world model of sound from action, 2026.

\bibitem{wu2026visreg}
Haiyu Wu, Randall Balestriero, and Morgan Levine.
\newblock Visreg: Variance-invariance-sketching regularization for jepa
  training, 2026.

\bibitem{wu2023melodyglm}
Xiao Wu, Zihao Huang, Kai Zhang, Jun Yu, Xu~Tan, Tao Zhang, Yan Li, Zhan Wang,
  and Lingling Sun.
\newblock {MelodyGLM}: Multi-task pre-training for symbolic melody generation.
\newblock {\em arXiv preprint arXiv:2309.10738}, 2023.

\bibitem{tan2026midillama}
Meng Yang, Jon McCormack, Maria~Teresa Llano, Wanchao Su, and Chao Lei.
\newblock {MIDI-LLaMA}: An instruction-following multimodal {LLM} for symbolic
  music understanding.
\newblock {\em arXiv preprint arXiv:2601.21740}, 2026.

\bibitem{anon2026arima}
Mingyang Yao and Zhaoxiang Feng.
\newblock {ARIMA}: Reconstruction-grounded predictive representation learning
  for symbolic music.
\newblock {\em arXiv preprint arXiv:2607.10003}, 2026.

\bibitem{zhou2025abceval}
Jiahao Zhao, Yunjia Li, Wei Li, and Kazuyoshi Yoshii.
\newblock {ABC-Eval}: Benchmarking large language models on symbolic music
  understanding and instruction following.
\newblock {\em arXiv preprint arXiv:2509.23350}, 2025.

\bibitem{zheng2025rae}
Bowen Zheng, Nan Ma, Shengbang Tong, and Saining Xie.
\newblock Diffusion transformers with representation autoencoders.
\newblock {\em arXiv preprint arXiv:2510.11690}, 2025.

\end{thebibliography}

\newpage
\appendix
% AUTOGENERATED by make_appendix.py — edit the generator, not this file
\setcounter{table}{0}\renewcommand{\thetable}{A\arabic{table}}

\section*{Supplemental Materials / Appendices}
The following appendices archive the complete measurement record behind the
paper. An interactive version of these tables, with listening examples and
the live demo, is maintained at
\href{https://drscotthawley.github.io/midi-rae-jepa-son/}{\nolinkurl{drscotthawley.github.io/midi-rae-jepa-son}}.

\section{Encoder training objective: equations and details}
\label{app:losses}
The total training objective combines four terms:
\begin{equation}
  \mathcal{L} = \lambda \mathcal{L}_{\text{equiv}} + (1-\lambda) \mathcal{L}_{\text{SIGReg}} + \lambda_{\text{MEP}} \mathcal{L}_{\text{MEP}} + \lambda_{\text{fact}} \mathcal{L}_{\text{fact}}
\end{equation}
applied level-wise: $\mathcal{L}_{\text{equiv}}$ and $\mathcal{L}_{\text{SIGReg}}$
act only at L0--L3 (at L4 and L5, SIGReg washes out the fine note-level
structure those levels carry, and patch-level equivariance is less meaningful
at those scales), and $\mathcal{L}_{\text{fact}}$, described at the end of this
appendix, acts only at the three coarsest levels L0--L2.

\paragraph{View generation.}
Two views are created by applying random shifts in time and pitch:
$\mathbf{x}_1$ is the original crop and $\mathbf{x}_2$ is shifted by
$(\Delta_x, \Delta_y)$ pixels. Both views are full $128\times128$ windows
extracted from the song's complete piano roll; a ``shift'' is implemented by
sampling the second window at a position offset by $(\Delta_x, \Delta_y)$, so
views always retain the encoder's input size.

\paragraph{Equivariance loss.}
A naive attraction loss collapses all shifted pairs to identical embeddings
regardless of shift magnitude. Instead we enforce a \emph{target} embedding
distance proportional to the shift magnitude, so that larger shifts produce
proportionally more distant embeddings:
\begin{equation}
  \mathcal{L}_{\text{equiv}}(\mathbf{z}_1, \mathbf{z}_2, \boldsymbol{\delta}) =
  \left(\|\mathbf{z}_1 - \mathbf{z}_2\| - \alpha \sqrt{d} \, \|\hat{\boldsymbol{\delta}}\|\right)^2
\end{equation}
where $\mathbf{z}_1$ is the teacher embedding of $\mathbf{x}_1$ and
$\mathbf{z}_2$ is the student embedding of $\mathbf{x}_2$ (see the EMA teacher
paragraph below), $\|\cdot\|$ denotes the Euclidean (L2) norm, $d$ is the
embedding dimension, $\hat{\boldsymbol{\delta}}$ is the shift
$\boldsymbol{\delta}=(\Delta_x,\Delta_y)$ with each component divided by its
per-axis maximum ($\Delta t_{\max}$, $\Delta p_{\max}$) so that each component
lies in $[0,1]$, and $\alpha$ is a scalar hyperparameter. The $\sqrt{d}$ factor
accounts for concentration of measure in high-dimensional spaces, giving
$\alpha$ consistent semantic meaning across all hierarchy levels. Unlike a
hinge loss, this smooth quadratic both attracts pairs that are too far apart
\emph{and} repels pairs that are too close. Shifts are sampled from a
$\mathrm{Beta}(2,2)$ distribution up to $(\Delta t_{\max}, \Delta p_{\max})$
pixels, with $\Delta p_{\max}=12$ (one octave) and $\Delta t_{\max}=48$ for the
models in this paper.

\paragraph{Chunked SIGReg.}
SIGReg uses the Epps--Pulley characteristic-function test to enforce an
isotropic Gaussian prior on embeddings~\cite{balestriero2025lejepa}. SIGReg
computes a tensor of shape $(N, N_{\text{slices}}, T)$ with
$N_{\text{slices}}=256$; on 16\,GB GPUs this tensor alone consumed
$\sim$5\,GB. We chunk the slice dimension into groups of 32 and accumulate in
float32:
\begin{equation}
  \mathcal{L}_{\text{SIGReg}} \approx \textstyle\sum_{k} \mathcal{L}^{(k)}_{\text{SIGReg}}
\end{equation}
where $k$ indexes the chunks. This reduces peak VRAM by $\sim$5\,GB and
counterintuitively improves throughput by reducing memory-allocator pressure.
SIGReg is applied only to student embeddings $\mathbf{z}_2$.

\paragraph{EMA teacher and masked embedding prediction.}
Following DINO~\cite{caron2021dino} and I-JEPA~\cite{assran2023ijepa}, we
maintain an EMA teacher
$\theta_T \leftarrow \eta\,\theta_T + (1-\eta)\,\theta_S$ with $\eta = 0.96$,
initialised from the student. The teacher produces stable target embeddings
$\mathbf{z}_1$ for the equivariance loss and processes $\mathbf{x}_2$ without
masking to supply prediction targets for the MEP; a lightweight predictor,
conditioned on the student's full-context representation, predicts the teacher
embeddings at randomly masked patch positions. The teacher is \emph{not}
redundant with SIGReg: SIGReg alone prevents collapse; the teacher exists to
provide stable, slowly-varying regression targets. The MEP loss is averaged
over hierarchy levels:
\begin{equation}
  \mathcal{L}_{\text{MEP}} = \frac{1}{L}\sum_{\ell=1}^{L}
    \left\|\hat{\mathbf{e}}_\ell - \mathbf{e}^T_\ell\right\|^2
\end{equation}
where $\mathbf{e}^T_\ell$ are the teacher embeddings and
$\hat{\mathbf{e}}_\ell$ the predictor's estimates at masked positions.

\begin{figure}[tb]
  \centering
  \includegraphics[width=0.8\columnwidth]{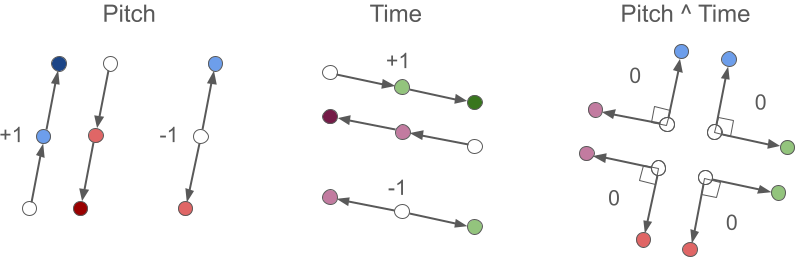}
  \caption{Soft factorization targets. Differences between pairs of embeddings
  are encouraged to be parallel ($t{=}+1$), anti-parallel ($t{=}-1$), or
  orthogonal ($t{=}0$) depending on augmentation type and sign. Directions are
  not prescribed; only their pairwise geometry is constrained.}
  \label{fig:app-softfact}
\end{figure}

\begin{figure}[tb]
  \centering
  \includegraphics[width=.8\columnwidth]{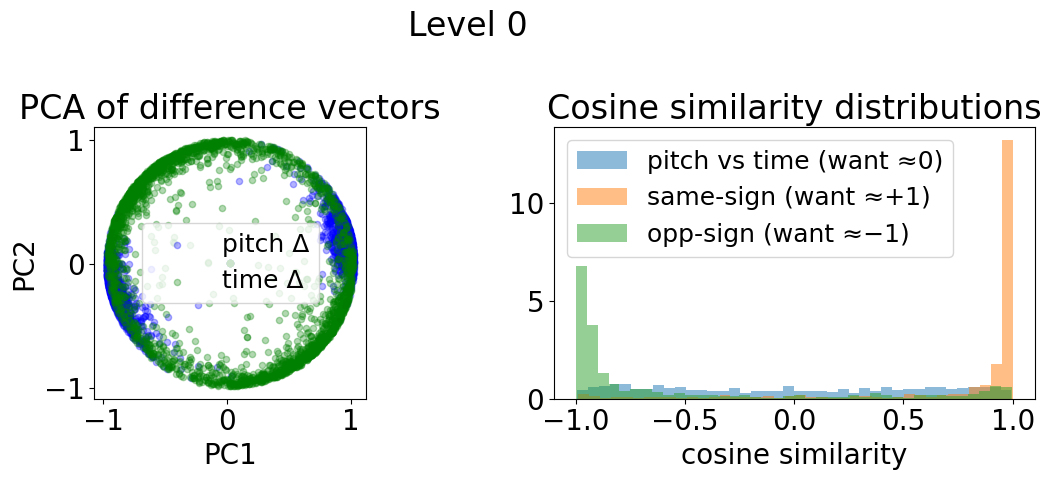}
  \caption{Example soft factorization results, Level 0. Left: PCA of normalized
  embedding difference vectors for the three target types, showing the expected
  geometric relationships. Right: histograms of cosine-similarity distributions
  for each pair type, with means indicated by dashed lines.}
  \label{fig:app-softfact-results}
\end{figure}

\begin{figure}[tb]
  \centering
  \includegraphics[width=.75\columnwidth]{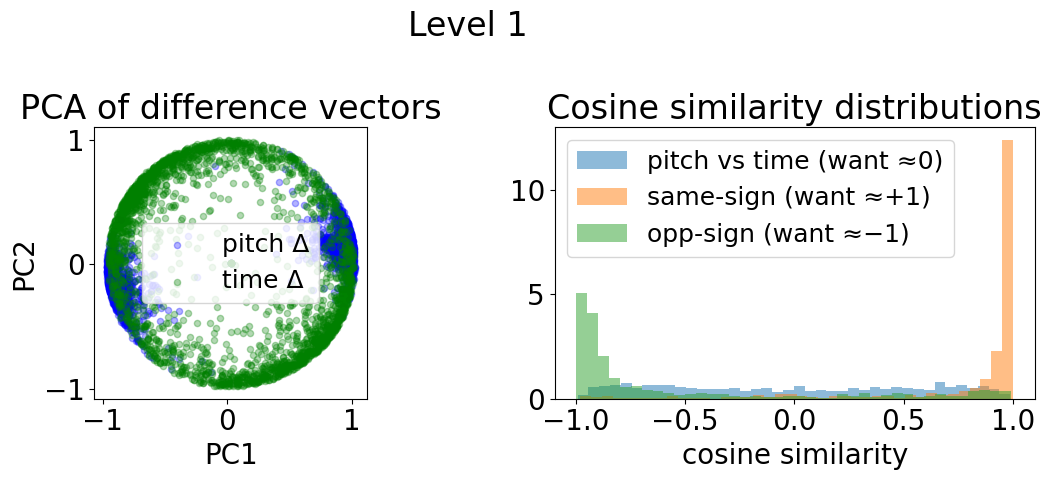}\\
  \includegraphics[width=.75\columnwidth]{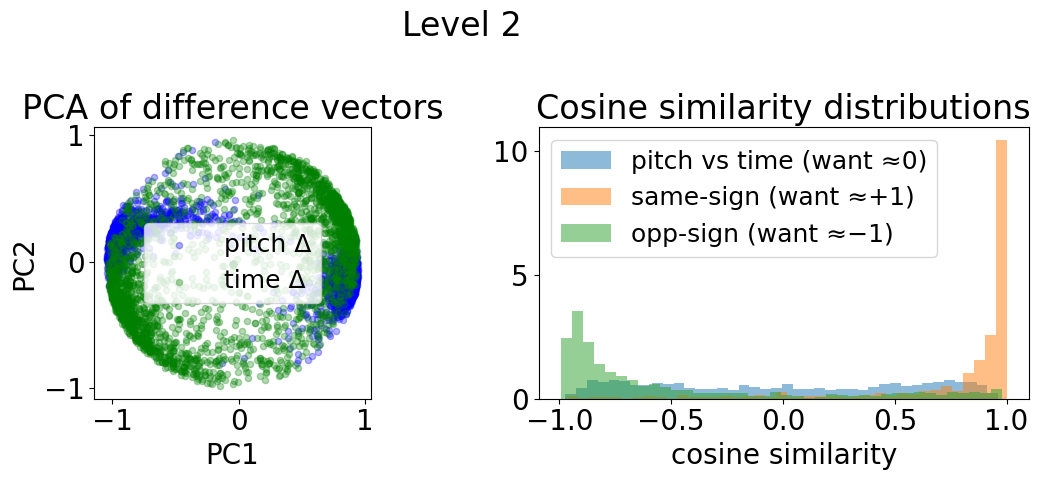}\\
  \includegraphics[width=.75\columnwidth]{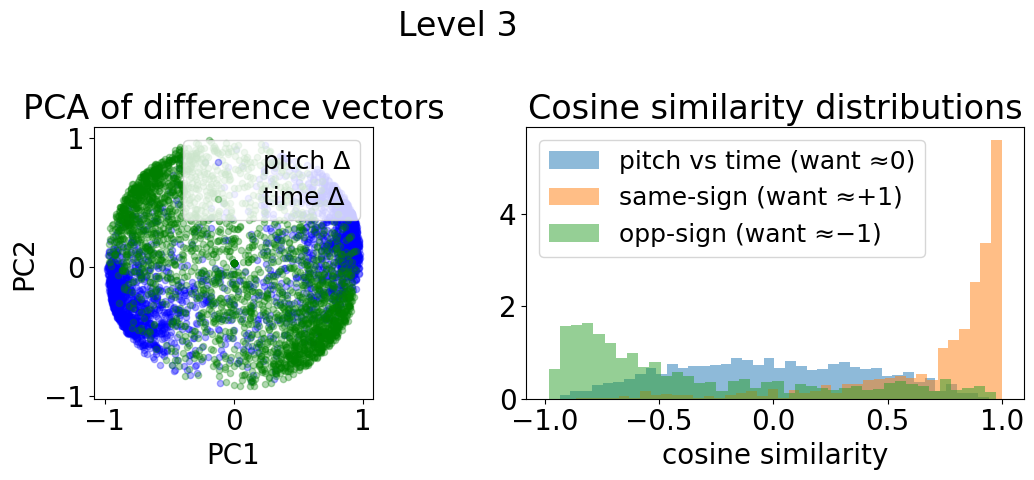}
  \caption{Soft factorization results for levels L1, L2, and L3 (top to
  bottom), in the format of Figure~\ref{fig:app-softfact-results}.}
  \label{fig:app-softfact-L023}
\end{figure}

\begin{figure}[tb]
  \centering
  \includegraphics[width=.8\columnwidth]{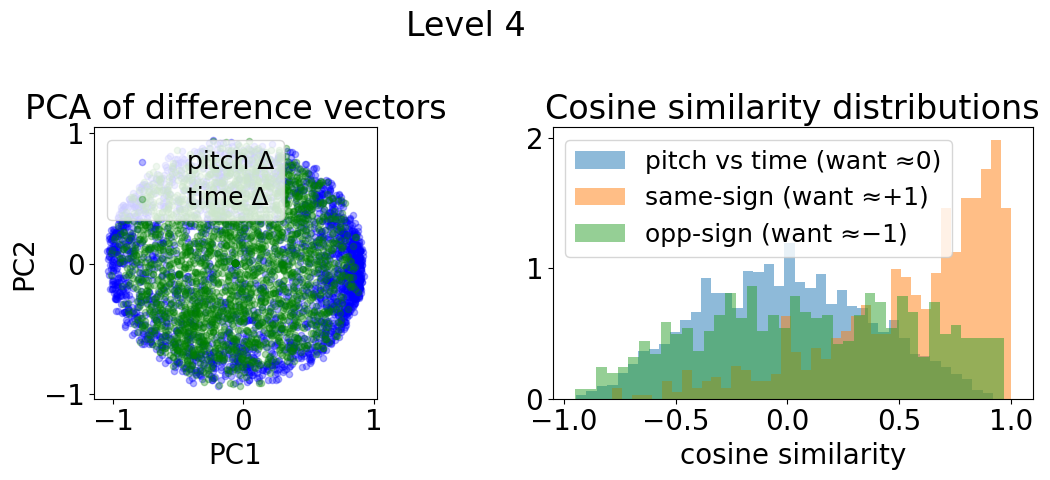}\\
  \includegraphics[width=.8\columnwidth]{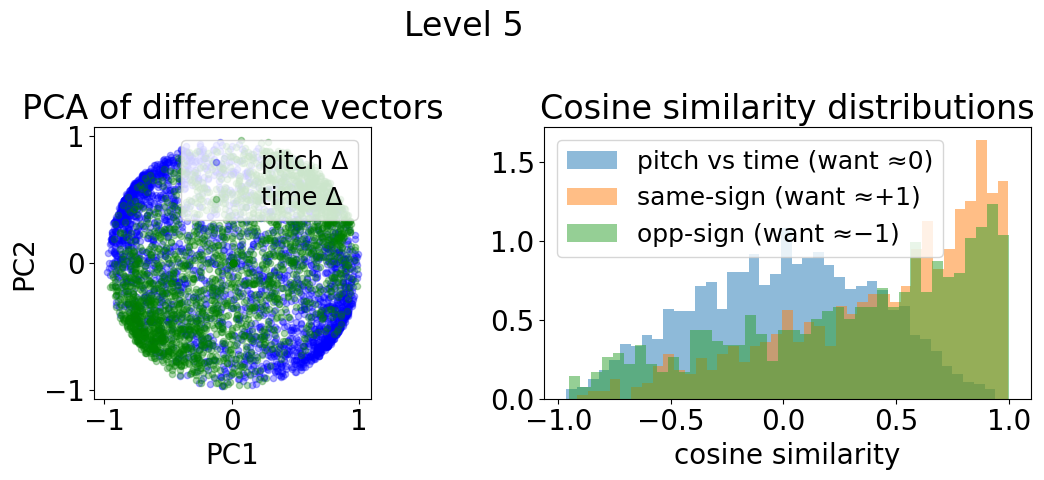}
  \caption{Soft factorization results for L4 and L5 (top to bottom). The
  factorization loss is applied only at L0--L2; the geometry visible at
  L3--L5 is inherited through the shared backbone rather than directly
  enforced.}
  \label{fig:app-softfact-L45}
\end{figure}

\paragraph{Soft factorization of pitch and time.}
Rather than enforcing pitch/time factorization architecturally---which prior
work has shown loses meaningful cross-dimension
interactions~\cite{wu2023melodyglm,chen2020musicsketchnet}---we impose a soft
geometric constraint through a cosine-similarity loss on augmentation
difference vectors. The loss is applied only at the three coarsest levels
(L0--L2). For a triplet of embeddings $\mathbf{z}_a$, $\mathbf{z}_1$,
$\mathbf{z}_2$, where the anchor $\mathbf{z}_a$ is the embedding of the
original unshifted crop and $\mathbf{z}_1$, $\mathbf{z}_2$ are embeddings of
two independently shifted crops of the same excerpt:
\begin{align}
  \mathbf{d}_1 &= \mathbf{z}_1 - \mathbf{z}_a, \quad \mathbf{d}_2 = \mathbf{z}_2 - \mathbf{z}_a \notag \\
  \mathcal{L}_{\text{fact}} &= \left( \cos(\mathbf{d}_1, \mathbf{d}_2) - t \right)^2
\end{align}
where the target $t \in \{+1, 0, -1\}$ encodes the geometric relationship
dictated by the augmentation types: same-type same-sign shifts should be
\emph{parallel} ($t{=}+1$), same-type opposite-sign shifts \emph{anti-parallel}
($t{=}-1$), and cross-type shifts (one pitch, one time) \emph{orthogonal}
($t{=}0$). No directions are prescribed; the system discovers them freely
subject to these pairwise constraints. Figure~\ref{fig:app-softfact}
illustrates the three target relationships, and
Figures~\ref{fig:app-softfact-results}--\ref{fig:app-softfact-L45} show the resulting geometry in a
trained encoder.

\section{Full STORMBIRD probe results}
\label{app:stormbird}
The probes of musical qualities used here form a suite of tests called
STORMBIRD\footnote{\href{https://github.com/drscotthawley/stormbird}{\nolinkurl{github.com/drscotthawley/stormbird}}}
(Sweeping Test Of Representational Music Benchmarks, Information Retrieval \&
Diagnostics).
Per-level results for the complete probe suite over all trained encoders
(Table~\ref{tab:app-roster}); the main text's consolidated table reports only
each model's best level. Values are reported to 3 decimals.

\begin{table}[H]
\centering
\small
\caption{Encoder roster.}
\label{tab:app-roster}
\setlength{\tabcolsep}{4pt}
\begin{tabular}{llp{7.4cm}}
\toprule
label & run tag & characteristics \\
\midrule
baseline & c55enc3\_vkrDxf & C55cUL recipe, no auxiliary supervision; 250 epochs, POP909. The unsupervised reference. \\
chord-0.5 & chord\_ynHbKs & + chord heads (root 12-way, quality 4-way) on coarse levels L0--L3, $\lambda$=0.5; 250 epochs. \\
chord-0.1 & chord01b & Chord heads L0--L3, $\lambda$=0.1; 250 epochs. \\
chord-1.0 & chord10 & Chord heads L0--L3, $\lambda$=1.0; 250 epochs. \\
phrase & phrase\_B2AWUj & + phrase-boundary head (1 logit/patch, Dai et al.\ annotations) on L0--L3, $\lambda$=0.5; 250 epochs. \\
chall-scr1 & mchordall\_screen & Chord heads on ALL levels L0--L5, $\lambda$=1.0 (0.167/level after averaging); 100-epoch screen. \\
chall-scr1.5 & mchordall\_matched & Chord heads on all levels, $\lambda$=1.5 = 0.25/level, weight-matched to the coarse-only $\lambda$=1.0 arm; 100-epoch screen. \\
lakh1x & lakh1x\_DKCVEJ & Baseline recipe on Lakh512\_1x (909 Lakh images); probes are cross-domain. \\
lakh4x & lakh4x\_qw3nqL & Same, on the nested 4$\times$ Lakh subset (3636 images). \\
\bottomrule
\end{tabular}
\end{table}

\begin{table}[H]
\centering
\small
\caption{Chord quality accuracy (chance $\approx$.50) $\uparrow$. Here and for all following tables, cells are shaded per row with the plasma colormap (yellow = best, honoring each caption's arrow), and the best value in each row is bold.}
\label{tab:app-chord_quality}
\setlength{\tabcolsep}{4pt}
\begin{tabular}{lccccccccc}
\toprule
Level & \rotatebox{55}{\scriptsize baseline} & \rotatebox{55}{\scriptsize chord-0.5} & \rotatebox{55}{\scriptsize chord-0.1} & \rotatebox{55}{\scriptsize chord-1.0} & \rotatebox{55}{\scriptsize phrase} & \rotatebox{55}{\scriptsize chall-scr1} & \rotatebox{55}{\scriptsize chall-scr1.5} & \rotatebox{55}{\scriptsize lakh1x} & \rotatebox{55}{\scriptsize lakh4x} \\
\midrule
L0 & \cellcolor[RGB]{208,77,115}\textcolor{black}{.527} & \cellcolor[RGB]{223,97,99}\textcolor{black}{.530} & \cellcolor[RGB]{203,71,119}\textcolor{black}{.526} & \cellcolor[RGB]{193,60,128}\textcolor{white}{.524} & \cellcolor[RGB]{139,9,164}\textcolor{white}{.516} & \cellcolor[RGB]{239,248,33}\textcolor{black}{\textbf{.549}} & \cellcolor[RGB]{12,7,134}\textcolor{white}{.502} & \cellcolor[RGB]{251,163,55}\textcolor{black}{.539} & \cellcolor[RGB]{239,126,78}\textcolor{black}{.534} \\
L1 & \cellcolor[RGB]{144,14,163}\textcolor{white}{.526} & \cellcolor[RGB]{98,0,166}\textcolor{white}{.522} & \cellcolor[RGB]{201,69,121}\textcolor{black}{.532} & \cellcolor[RGB]{248,150,63}\textcolor{black}{.540} & \cellcolor[RGB]{239,248,33}\textcolor{black}{\textbf{.548}} & \cellcolor[RGB]{228,106,93}\textcolor{black}{.536} & \cellcolor[RGB]{12,7,134}\textcolor{white}{.516} & \cellcolor[RGB]{251,211,36}\textcolor{black}{.545} & \cellcolor[RGB]{240,128,77}\textcolor{black}{.538} \\
L2 & \cellcolor[RGB]{242,132,75}\textcolor{black}{.540} & \cellcolor[RGB]{12,7,134}\textcolor{white}{.513} & \cellcolor[RGB]{226,102,96}\textcolor{black}{.536} & \cellcolor[RGB]{250,213,36}\textcolor{black}{.548} & \cellcolor[RGB]{86,1,163}\textcolor{white}{.519} & \cellcolor[RGB]{239,248,33}\textcolor{black}{\textbf{.551}} & \cellcolor[RGB]{252,173,49}\textcolor{black}{.544} & \cellcolor[RGB]{246,229,37}\textcolor{black}{.549} & \cellcolor[RGB]{214,85,109}\textcolor{black}{.534} \\
L3 & \cellcolor[RGB]{253,187,43}\textcolor{black}{.540} & \cellcolor[RGB]{251,208,36}\textcolor{black}{.543} & \cellcolor[RGB]{189,55,132}\textcolor{white}{.517} & \cellcolor[RGB]{12,7,134}\textcolor{white}{.492} & \cellcolor[RGB]{149,17,161}\textcolor{white}{.510} & \cellcolor[RGB]{250,216,36}\textcolor{black}{.544} & \cellcolor[RGB]{247,147,65}\textcolor{black}{.534} & \cellcolor[RGB]{239,248,33}\textcolor{black}{\textbf{.548}} & \cellcolor[RGB]{233,114,87}\textcolor{black}{.528} \\
L4 & \cellcolor[RGB]{209,79,113}\textcolor{black}{.540} & \cellcolor[RGB]{164,30,152}\textcolor{white}{.534} & \cellcolor[RGB]{86,1,163}\textcolor{white}{.528} & \cellcolor[RGB]{12,7,134}\textcolor{white}{.523} & \cellcolor[RGB]{98,0,166}\textcolor{white}{.529} & \cellcolor[RGB]{239,248,33}\textcolor{black}{\textbf{.555}} & \cellcolor[RGB]{209,79,113}\textcolor{black}{.540} & \cellcolor[RGB]{164,30,152}\textcolor{white}{.534} & \cellcolor[RGB]{149,17,161}\textcolor{white}{.533} \\
L5 & \cellcolor[RGB]{199,66,123}\textcolor{white}{.536} & \cellcolor[RGB]{240,246,35}\textcolor{black}{.549} & \cellcolor[RGB]{51,4,151}\textcolor{white}{.526} & \cellcolor[RGB]{12,7,134}\textcolor{white}{.524} & \cellcolor[RGB]{204,72,118}\textcolor{black}{.537} & \cellcolor[RGB]{239,248,33}\textcolor{black}{\textbf{.549}} & \cellcolor[RGB]{153,20,159}\textcolor{white}{.532} & \cellcolor[RGB]{237,123,81}\textcolor{black}{.541} & \cellcolor[RGB]{252,201,38}\textcolor{black}{.546} \\
\bottomrule
\end{tabular}
\end{table}

\begin{table}[H]
\centering
\small
\caption{Root note accuracy (12-class, chance .083) $\uparrow$}
\label{tab:app-root_note}
\setlength{\tabcolsep}{4pt}
\begin{tabular}{lccccccccc}
\toprule
Level & \rotatebox{55}{\scriptsize baseline} & \rotatebox{55}{\scriptsize chord-0.5} & \rotatebox{55}{\scriptsize chord-0.1} & \rotatebox{55}{\scriptsize chord-1.0} & \rotatebox{55}{\scriptsize phrase} & \rotatebox{55}{\scriptsize chall-scr1} & \rotatebox{55}{\scriptsize chall-scr1.5} & \rotatebox{55}{\scriptsize lakh1x} & \rotatebox{55}{\scriptsize lakh4x} \\
\midrule
L0 & \cellcolor[RGB]{27,6,140}\textcolor{white}{.094} & \cellcolor[RGB]{247,224,36}\textcolor{black}{.457} & \cellcolor[RGB]{19,6,137}\textcolor{white}{.090} & \cellcolor[RGB]{239,248,33}\textcolor{black}{\textbf{.480}} & \cellcolor[RGB]{12,7,134}\textcolor{white}{.086} & \cellcolor[RGB]{165,31,151}\textcolor{white}{.229} & \cellcolor[RGB]{253,182,45}\textcolor{black}{.418} & \cellcolor[RGB]{35,5,144}\textcolor{white}{.101} & \cellcolor[RGB]{41,5,147}\textcolor{white}{.106} \\
L1 & \cellcolor[RGB]{12,7,134}\textcolor{white}{.090} & \cellcolor[RGB]{239,248,33}\textcolor{black}{\textbf{.570}} & \cellcolor[RGB]{29,6,141}\textcolor{white}{.102} & \cellcolor[RGB]{243,236,38}\textcolor{black}{.555} & \cellcolor[RGB]{29,6,141}\textcolor{white}{.102} & \cellcolor[RGB]{194,61,128}\textcolor{white}{.313} & \cellcolor[RGB]{248,223,36}\textcolor{black}{.541} & \cellcolor[RGB]{39,5,146}\textcolor{white}{.111} & \cellcolor[RGB]{31,5,142}\textcolor{white}{.104} \\
L2 & \cellcolor[RGB]{12,7,134}\textcolor{white}{.107} & \cellcolor[RGB]{243,238,38}\textcolor{black}{.594} & \cellcolor[RGB]{49,4,150}\textcolor{white}{.138} & \cellcolor[RGB]{239,248,33}\textcolor{black}{\textbf{.607}} & \cellcolor[RGB]{27,6,140}\textcolor{white}{.118} & \cellcolor[RGB]{189,55,132}\textcolor{white}{.330} & \cellcolor[RGB]{248,223,36}\textcolor{black}{.578} & \cellcolor[RGB]{41,5,147}\textcolor{white}{.132} & \cellcolor[RGB]{16,7,135}\textcolor{white}{.109} \\
L3 & \cellcolor[RGB]{12,7,134}\textcolor{white}{.132} & \cellcolor[RGB]{251,209,36}\textcolor{black}{.582} & \cellcolor[RGB]{90,0,165}\textcolor{white}{.212} & \cellcolor[RGB]{239,248,33}\textcolor{black}{\textbf{.627}} & \cellcolor[RGB]{45,4,148}\textcolor{white}{.160} & \cellcolor[RGB]{175,40,144}\textcolor{white}{.328} & \cellcolor[RGB]{253,195,40}\textcolor{black}{.564} & \cellcolor[RGB]{54,4,152}\textcolor{white}{.169} & \cellcolor[RGB]{47,4,149}\textcolor{white}{.161} \\
L4 & \cellcolor[RGB]{21,6,138}\textcolor{white}{.203} & \cellcolor[RGB]{218,90,104}\textcolor{black}{.405} & \cellcolor[RGB]{78,2,161}\textcolor{white}{.246} & \cellcolor[RGB]{236,120,83}\textcolor{black}{.442} & \cellcolor[RGB]{12,7,134}\textcolor{white}{.199} & \cellcolor[RGB]{142,12,164}\textcolor{white}{.305} & \cellcolor[RGB]{239,248,33}\textcolor{black}{\textbf{.563}} & \cellcolor[RGB]{27,6,140}\textcolor{white}{.206} & \cellcolor[RGB]{12,7,134}\textcolor{white}{.198} \\
L5 & \cellcolor[RGB]{212,82,110}\textcolor{black}{.243} & \cellcolor[RGB]{205,73,117}\textcolor{black}{.242} & \cellcolor[RGB]{250,157,58}\textcolor{black}{.253} & \cellcolor[RGB]{243,135,72}\textcolor{black}{.251} & \cellcolor[RGB]{29,6,141}\textcolor{white}{.221} & \cellcolor[RGB]{203,71,119}\textcolor{black}{.242} & \cellcolor[RGB]{251,163,55}\textcolor{black}{.254} & \cellcolor[RGB]{12,7,134}\textcolor{white}{.220} & \cellcolor[RGB]{239,248,33}\textcolor{black}{\textbf{.263}} \\
\bottomrule
\end{tabular}
\end{table}

\begin{table}[H]
\centering
\small
\caption{Chroma regression $R^2$ $\uparrow$}
\label{tab:app-chroma_r2}
\setlength{\tabcolsep}{4pt}
\begin{tabular}{lccccccccc}
\toprule
Level & \rotatebox{55}{\scriptsize baseline} & \rotatebox{55}{\scriptsize chord-0.5} & \rotatebox{55}{\scriptsize chord-0.1} & \rotatebox{55}{\scriptsize chord-1.0} & \rotatebox{55}{\scriptsize phrase} & \rotatebox{55}{\scriptsize chall-scr1} & \rotatebox{55}{\scriptsize chall-scr1.5} & \rotatebox{55}{\scriptsize lakh1x} & \rotatebox{55}{\scriptsize lakh4x} \\
\midrule
L0 & \cellcolor[RGB]{12,7,134}\textcolor{white}{$-$.019} & \cellcolor[RGB]{243,238,38}\textcolor{black}{.686} & \cellcolor[RGB]{12,7,134}\textcolor{white}{$-$.020} & \cellcolor[RGB]{239,248,33}\textcolor{black}{\textbf{.704}} & \cellcolor[RGB]{12,7,134}\textcolor{white}{$-$.011} & \cellcolor[RGB]{132,5,166}\textcolor{white}{.188} & \cellcolor[RGB]{245,231,38}\textcolor{black}{.676} & \cellcolor[RGB]{12,7,134}\textcolor{white}{$-$.006} & \cellcolor[RGB]{12,7,134}\textcolor{white}{$-$.016} \\
L1 & \cellcolor[RGB]{12,7,134}\textcolor{white}{$-$.019} & \cellcolor[RGB]{247,224,36}\textcolor{black}{.692} & \cellcolor[RGB]{27,6,140}\textcolor{white}{.016} & \cellcolor[RGB]{239,248,33}\textcolor{black}{\textbf{.733}} & \cellcolor[RGB]{29,6,141}\textcolor{white}{.019} & \cellcolor[RGB]{178,44,142}\textcolor{white}{.300} & \cellcolor[RGB]{243,236,38}\textcolor{black}{.711} & \cellcolor[RGB]{39,5,146}\textcolor{white}{.032} & \cellcolor[RGB]{16,7,135}\textcolor{white}{.004} \\
L2 & \cellcolor[RGB]{12,7,134}\textcolor{white}{.064} & \cellcolor[RGB]{245,231,38}\textcolor{black}{.731} & \cellcolor[RGB]{64,3,156}\textcolor{white}{.133} & \cellcolor[RGB]{244,234,38}\textcolor{black}{.737} & \cellcolor[RGB]{39,5,146}\textcolor{white}{.096} & \cellcolor[RGB]{213,83,109}\textcolor{black}{.444} & \cellcolor[RGB]{239,248,33}\textcolor{black}{\textbf{.760}} & \cellcolor[RGB]{64,3,156}\textcolor{white}{.132} & \cellcolor[RGB]{21,6,138}\textcolor{white}{.073} \\
L3 & \cellcolor[RGB]{12,7,134}\textcolor{white}{.277} & \cellcolor[RGB]{239,248,33}\textcolor{black}{\textbf{.826}} & \cellcolor[RGB]{100,0,167}\textcolor{white}{.380} & \cellcolor[RGB]{239,248,33}\textcolor{black}{.825} & \cellcolor[RGB]{21,6,138}\textcolor{white}{.286} & \cellcolor[RGB]{208,77,115}\textcolor{black}{.563} & \cellcolor[RGB]{240,246,35}\textcolor{black}{.823} & \cellcolor[RGB]{112,0,168}\textcolor{white}{.396} & \cellcolor[RGB]{33,5,143}\textcolor{white}{.296} \\
L4 & \cellcolor[RGB]{12,7,134}\textcolor{white}{.310} & \cellcolor[RGB]{241,243,38}\textcolor{black}{.828} & \cellcolor[RGB]{79,2,162}\textcolor{white}{.380} & \cellcolor[RGB]{239,248,33}\textcolor{black}{\textbf{.835}} & \cellcolor[RGB]{170,36,148}\textcolor{white}{.509} & \cellcolor[RGB]{228,106,93}\textcolor{black}{.635} & \cellcolor[RGB]{242,240,38}\textcolor{black}{.824} & \cellcolor[RGB]{223,97,99}\textcolor{black}{.621} & \cellcolor[RGB]{112,0,168}\textcolor{white}{.424} \\
L5 & \cellcolor[RGB]{200,68,122}\textcolor{white}{.646} & \cellcolor[RGB]{245,231,38}\textcolor{black}{.824} & \cellcolor[RGB]{12,7,134}\textcolor{white}{.460} & \cellcolor[RGB]{239,248,33}\textcolor{black}{\textbf{.839}} & \cellcolor[RGB]{249,154,60}\textcolor{black}{.751} & \cellcolor[RGB]{242,132,75}\textcolor{black}{.726} & \cellcolor[RGB]{249,219,36}\textcolor{black}{.813} & \cellcolor[RGB]{251,162,56}\textcolor{black}{.758} & \cellcolor[RGB]{177,43,143}\textcolor{white}{.614} \\
\bottomrule
\end{tabular}
\end{table}

\begin{table}[H]
\centering
\small
\caption{Key detection accuracy (24-class) $\uparrow$}
\label{tab:app-key_detection}
\setlength{\tabcolsep}{4pt}
\begin{tabular}{lccccccccc}
\toprule
Level & \rotatebox{55}{\scriptsize baseline} & \rotatebox{55}{\scriptsize chord-0.5} & \rotatebox{55}{\scriptsize chord-0.1} & \rotatebox{55}{\scriptsize chord-1.0} & \rotatebox{55}{\scriptsize phrase} & \rotatebox{55}{\scriptsize chall-scr1} & \rotatebox{55}{\scriptsize chall-scr1.5} & \rotatebox{55}{\scriptsize lakh1x} & \rotatebox{55}{\scriptsize lakh4x} \\
\midrule
L0 & \cellcolor[RGB]{16,7,135}\textcolor{white}{.043} & \cellcolor[RGB]{239,248,33}\textcolor{black}{\textbf{.415}} & \cellcolor[RGB]{12,7,134}\textcolor{white}{.041} & \cellcolor[RGB]{247,226,37}\textcolor{black}{.396} & \cellcolor[RGB]{31,5,142}\textcolor{white}{.052} & \cellcolor[RGB]{144,14,163}\textcolor{white}{.154} & \cellcolor[RGB]{253,175,49}\textcolor{black}{.349} & \cellcolor[RGB]{33,5,143}\textcolor{white}{.054} & \cellcolor[RGB]{33,5,143}\textcolor{white}{.053} \\
L1 & \cellcolor[RGB]{12,7,134}\textcolor{white}{.047} & \cellcolor[RGB]{239,248,33}\textcolor{black}{\textbf{.602}} & \cellcolor[RGB]{12,7,134}\textcolor{white}{.047} & \cellcolor[RGB]{247,226,37}\textcolor{black}{.572} & \cellcolor[RGB]{12,7,134}\textcolor{white}{.047} & \cellcolor[RGB]{153,20,159}\textcolor{white}{.228} & \cellcolor[RGB]{246,228,37}\textcolor{black}{.575} & \cellcolor[RGB]{33,5,143}\textcolor{white}{.065} & \cellcolor[RGB]{21,6,138}\textcolor{white}{.055} \\
L2 & \cellcolor[RGB]{12,7,134}\textcolor{white}{.054} & \cellcolor[RGB]{243,238,38}\textcolor{black}{.688} & \cellcolor[RGB]{27,6,140}\textcolor{white}{.068} & \cellcolor[RGB]{239,248,33}\textcolor{black}{\textbf{.706}} & \cellcolor[RGB]{31,5,142}\textcolor{white}{.073} & \cellcolor[RGB]{158,25,156}\textcolor{white}{.277} & \cellcolor[RGB]{245,233,38}\textcolor{black}{.681} & \cellcolor[RGB]{33,5,143}\textcolor{white}{.075} & \cellcolor[RGB]{21,6,138}\textcolor{white}{.062} \\
L3 & \cellcolor[RGB]{12,7,134}\textcolor{white}{.077} & \cellcolor[RGB]{248,223,36}\textcolor{black}{.698} & \cellcolor[RGB]{43,5,148}\textcolor{white}{.113} & \cellcolor[RGB]{239,248,33}\textcolor{black}{\textbf{.737}} & \cellcolor[RGB]{31,5,142}\textcolor{white}{.097} & \cellcolor[RGB]{133,6,166}\textcolor{white}{.256} & \cellcolor[RGB]{251,208,36}\textcolor{black}{.674} & \cellcolor[RGB]{24,6,139}\textcolor{white}{.089} & \cellcolor[RGB]{43,5,148}\textcolor{white}{.111} \\
L4 & \cellcolor[RGB]{12,7,134}\textcolor{white}{.128} & \cellcolor[RGB]{217,89,105}\textcolor{black}{.403} & \cellcolor[RGB]{35,5,144}\textcolor{white}{.146} & \cellcolor[RGB]{222,96,100}\textcolor{black}{.415} & \cellcolor[RGB]{21,6,138}\textcolor{white}{.134} & \cellcolor[RGB]{111,0,168}\textcolor{white}{.232} & \cellcolor[RGB]{239,248,33}\textcolor{black}{\textbf{.615}} & \cellcolor[RGB]{21,6,138}\textcolor{white}{.135} & \cellcolor[RGB]{21,6,138}\textcolor{white}{.134} \\
L5 & \cellcolor[RGB]{104,0,167}\textcolor{white}{.158} & \cellcolor[RGB]{228,106,93}\textcolor{black}{.174} & \cellcolor[RGB]{218,90,104}\textcolor{black}{.172} & \cellcolor[RGB]{239,248,33}\textcolor{black}{\textbf{.189}} & \cellcolor[RGB]{209,79,113}\textcolor{black}{.171} & \cellcolor[RGB]{223,97,99}\textcolor{black}{.173} & \cellcolor[RGB]{252,204,37}\textcolor{black}{.185} & \cellcolor[RGB]{12,7,134}\textcolor{white}{.151} & \cellcolor[RGB]{243,135,72}\textcolor{black}{.178} \\
\bottomrule
\end{tabular}
\end{table}

\begin{table}[H]
\centering
\small
\caption{Note density $R^2$ $\uparrow$}
\label{tab:app-density_r2}
\setlength{\tabcolsep}{4pt}
\begin{tabular}{lccccccccc}
\toprule
Level & \rotatebox{55}{\scriptsize baseline} & \rotatebox{55}{\scriptsize chord-0.5} & \rotatebox{55}{\scriptsize chord-0.1} & \rotatebox{55}{\scriptsize chord-1.0} & \rotatebox{55}{\scriptsize phrase} & \rotatebox{55}{\scriptsize chall-scr1} & \rotatebox{55}{\scriptsize chall-scr1.5} & \rotatebox{55}{\scriptsize lakh1x} & \rotatebox{55}{\scriptsize lakh4x} \\
\midrule
L0 & \cellcolor[RGB]{12,7,134}\textcolor{white}{.413} & \cellcolor[RGB]{231,110,90}\textcolor{black}{.584} & \cellcolor[RGB]{12,7,134}\textcolor{white}{.414} & \cellcolor[RGB]{239,125,79}\textcolor{black}{.597} & \cellcolor[RGB]{59,3,154}\textcolor{white}{.437} & \cellcolor[RGB]{249,154,60}\textcolor{black}{.620} & \cellcolor[RGB]{239,248,33}\textcolor{black}{\textbf{.683}} & \cellcolor[RGB]{252,169,52}\textcolor{black}{.631} & \cellcolor[RGB]{182,47,139}\textcolor{white}{.527} \\
L1 & \cellcolor[RGB]{12,7,134}\textcolor{white}{.508} & \cellcolor[RGB]{253,187,43}\textcolor{black}{.655} & \cellcolor[RGB]{64,3,156}\textcolor{white}{.525} & \cellcolor[RGB]{253,192,41}\textcolor{black}{.656} & \cellcolor[RGB]{115,0,168}\textcolor{white}{.546} & \cellcolor[RGB]{252,199,38}\textcolor{black}{.660} & \cellcolor[RGB]{239,248,33}\textcolor{black}{\textbf{.680}} & \cellcolor[RGB]{245,231,38}\textcolor{black}{.673} & \cellcolor[RGB]{201,69,121}\textcolor{black}{.593} \\
L2 & \cellcolor[RGB]{66,3,157}\textcolor{white}{.611} & \cellcolor[RGB]{252,203,37}\textcolor{black}{.745} & \cellcolor[RGB]{12,7,134}\textcolor{white}{.593} & \cellcolor[RGB]{252,172,50}\textcolor{black}{.732} & \cellcolor[RGB]{82,1,163}\textcolor{white}{.617} & \cellcolor[RGB]{240,246,35}\textcolor{black}{.762} & \cellcolor[RGB]{242,241,38}\textcolor{black}{.760} & \cellcolor[RGB]{239,248,33}\textcolor{black}{\textbf{.763}} & \cellcolor[RGB]{252,169,52}\textcolor{black}{.730} \\
L3 & \cellcolor[RGB]{137,8,165}\textcolor{white}{.723} & \cellcolor[RGB]{250,156,59}\textcolor{black}{.803} & \cellcolor[RGB]{12,7,134}\textcolor{white}{.676} & \cellcolor[RGB]{212,82,110}\textcolor{black}{.765} & \cellcolor[RGB]{174,39,145}\textcolor{white}{.740} & \cellcolor[RGB]{252,167,53}\textcolor{black}{.808} & \cellcolor[RGB]{239,248,33}\textcolor{black}{\textbf{.840}} & \cellcolor[RGB]{246,142,68}\textcolor{black}{.796} & \cellcolor[RGB]{237,123,81}\textcolor{black}{.786} \\
L4 & \cellcolor[RGB]{71,2,159}\textcolor{white}{.769} & \cellcolor[RGB]{239,248,33}\textcolor{black}{\textbf{.889}} & \cellcolor[RGB]{12,7,134}\textcolor{white}{.753} & \cellcolor[RGB]{253,187,43}\textcolor{black}{.869} & \cellcolor[RGB]{187,53,134}\textcolor{white}{.813} & \cellcolor[RGB]{215,86,108}\textcolor{black}{.829} & \cellcolor[RGB]{236,120,83}\textcolor{black}{.844} & \cellcolor[RGB]{248,223,36}\textcolor{black}{.881} & \cellcolor[RGB]{253,178,47}\textcolor{black}{.866} \\
L5 & \cellcolor[RGB]{233,114,87}\textcolor{black}{.884} & \cellcolor[RGB]{253,179,46}\textcolor{black}{.913} & \cellcolor[RGB]{12,7,134}\textcolor{white}{.787} & \cellcolor[RGB]{233,114,87}\textcolor{black}{.885} & \cellcolor[RGB]{239,125,79}\textcolor{black}{.890} & \cellcolor[RGB]{250,213,36}\textcolor{black}{.925} & \cellcolor[RGB]{164,30,152}\textcolor{white}{.841} & \cellcolor[RGB]{242,241,38}\textcolor{black}{.936} & \cellcolor[RGB]{239,248,33}\textcolor{black}{\textbf{.938}} \\
\bottomrule
\end{tabular}
\end{table}

\begin{table}[H]
\centering
\small
\caption{Cross-song distance ratio ($\downarrow$ lower is better)}
\label{tab:app-cross_song_ratio}
\setlength{\tabcolsep}{4pt}
\begin{tabular}{lccccccccc}
\toprule
Level & \rotatebox{55}{\scriptsize baseline} & \rotatebox{55}{\scriptsize chord-0.5} & \rotatebox{55}{\scriptsize chord-0.1} & \rotatebox{55}{\scriptsize chord-1.0} & \rotatebox{55}{\scriptsize phrase} & \rotatebox{55}{\scriptsize chall-scr1} & \rotatebox{55}{\scriptsize chall-scr1.5} & \rotatebox{55}{\scriptsize lakh1x} & \rotatebox{55}{\scriptsize lakh4x} \\
\midrule
L0 & \cellcolor[RGB]{59,3,154}\textcolor{white}{.803} & \cellcolor[RGB]{160,27,155}\textcolor{white}{.745} & \cellcolor[RGB]{178,44,142}\textcolor{white}{.731} & \cellcolor[RGB]{183,48,138}\textcolor{white}{.728} & \cellcolor[RGB]{239,248,33}\textcolor{black}{\textbf{.600}} & \cellcolor[RGB]{199,66,123}\textcolor{white}{.714} & \cellcolor[RGB]{253,181,45}\textcolor{black}{.636} & \cellcolor[RGB]{94,0,165}\textcolor{white}{.785} & \cellcolor[RGB]{12,7,134}\textcolor{white}{.822} \\
L1 & \cellcolor[RGB]{49,4,150}\textcolor{white}{.814} & \cellcolor[RGB]{180,45,141}\textcolor{white}{.729} & \cellcolor[RGB]{189,55,132}\textcolor{white}{.721} & \cellcolor[RGB]{173,38,146}\textcolor{white}{.736} & \cellcolor[RGB]{239,248,33}\textcolor{black}{\textbf{.586}} & \cellcolor[RGB]{143,13,163}\textcolor{white}{.757} & \cellcolor[RGB]{154,21,158}\textcolor{white}{.750} & \cellcolor[RGB]{160,27,155}\textcolor{white}{.745} & \cellcolor[RGB]{12,7,134}\textcolor{white}{.830} \\
L2 & \cellcolor[RGB]{12,7,134}\textcolor{white}{.848} & \cellcolor[RGB]{185,51,136}\textcolor{white}{.745} & \cellcolor[RGB]{200,68,122}\textcolor{white}{.731} & \cellcolor[RGB]{173,38,146}\textcolor{white}{.755} & \cellcolor[RGB]{239,248,33}\textcolor{black}{\textbf{.610}} & \cellcolor[RGB]{147,16,161}\textcolor{white}{.774} & \cellcolor[RGB]{147,16,161}\textcolor{white}{.774} & \cellcolor[RGB]{180,45,141}\textcolor{white}{.749} & \cellcolor[RGB]{56,4,153}\textcolor{white}{.829} \\
L3 & \cellcolor[RGB]{97,0,166}\textcolor{white}{.799} & \cellcolor[RGB]{39,5,146}\textcolor{white}{.837} & \cellcolor[RGB]{188,54,133}\textcolor{white}{.724} & \cellcolor[RGB]{114,0,168}\textcolor{white}{.787} & \cellcolor[RGB]{239,248,33}\textcolor{black}{\textbf{.566}} & \cellcolor[RGB]{140,10,164}\textcolor{white}{.767} & \cellcolor[RGB]{118,1,168}\textcolor{white}{.783} & \cellcolor[RGB]{241,130,76}\textcolor{black}{.651} & \cellcolor[RGB]{12,7,134}\textcolor{white}{.849} \\
L4 & \cellcolor[RGB]{169,35,149}\textcolor{white}{.847} & \cellcolor[RGB]{12,7,134}\textcolor{white}{.993} & \cellcolor[RGB]{248,150,63}\textcolor{black}{.700} & \cellcolor[RGB]{238,124,80}\textcolor{black}{.731} & \cellcolor[RGB]{239,248,33}\textcolor{black}{\textbf{.605}} & \cellcolor[RGB]{157,24,157}\textcolor{white}{.862} & \cellcolor[RGB]{82,1,163}\textcolor{white}{.938} & \cellcolor[RGB]{244,234,38}\textcolor{black}{.619} & \cellcolor[RGB]{193,60,128}\textcolor{white}{.814} \\
L5 & \cellcolor[RGB]{154,21,158}\textcolor{white}{.925} & \cellcolor[RGB]{12,7,134}\textcolor{white}{1.062} & \cellcolor[RGB]{246,145,66}\textcolor{black}{.755} & \cellcolor[RGB]{252,166,53}\textcolor{black}{.731} & \cellcolor[RGB]{239,248,33}\textcolor{black}{\textbf{.648}} & \cellcolor[RGB]{185,51,136}\textcolor{white}{.884} & \cellcolor[RGB]{157,24,157}\textcolor{white}{.921} & \cellcolor[RGB]{249,219,36}\textcolor{black}{.676} & \cellcolor[RGB]{242,133,74}\textcolor{black}{.770} \\
\bottomrule
\end{tabular}
\end{table}

\begin{table}[H]
\centering
\small
\caption{Time-translation $R^2$ $\uparrow$}
\label{tab:app-temporal_dist_r2}
\setlength{\tabcolsep}{4pt}
\begin{tabular}{lccccccccc}
\toprule
Level & \rotatebox{55}{\scriptsize baseline} & \rotatebox{55}{\scriptsize chord-0.5} & \rotatebox{55}{\scriptsize chord-0.1} & \rotatebox{55}{\scriptsize chord-1.0} & \rotatebox{55}{\scriptsize phrase} & \rotatebox{55}{\scriptsize chall-scr1} & \rotatebox{55}{\scriptsize chall-scr1.5} & \rotatebox{55}{\scriptsize lakh1x} & \rotatebox{55}{\scriptsize lakh4x} \\
\midrule
L0 & \cellcolor[RGB]{213,83,109}\textcolor{black}{.194} & \cellcolor[RGB]{202,70,120}\textcolor{black}{.192} & \cellcolor[RGB]{109,0,168}\textcolor{white}{.180} & \cellcolor[RGB]{114,0,168}\textcolor{white}{.180} & \cellcolor[RGB]{130,4,167}\textcolor{white}{.182} & \cellcolor[RGB]{12,7,134}\textcolor{white}{.171} & \cellcolor[RGB]{160,27,155}\textcolor{white}{.186} & \cellcolor[RGB]{239,248,33}\textcolor{black}{\textbf{.212}} & \cellcolor[RGB]{114,0,168}\textcolor{white}{.181} \\
L1 & \cellcolor[RGB]{239,248,33}\textcolor{black}{\textbf{.248}} & \cellcolor[RGB]{252,203,37}\textcolor{black}{.242} & \cellcolor[RGB]{237,123,81}\textcolor{black}{.230} & \cellcolor[RGB]{249,154,60}\textcolor{black}{.235} & \cellcolor[RGB]{253,185,43}\textcolor{black}{.240} & \cellcolor[RGB]{12,7,134}\textcolor{white}{.194} & \cellcolor[RGB]{95,0,166}\textcolor{white}{.203} & \cellcolor[RGB]{252,199,38}\textcolor{black}{.241} & \cellcolor[RGB]{92,0,165}\textcolor{white}{.203} \\
L2 & \cellcolor[RGB]{239,248,33}\textcolor{black}{\textbf{.250}} & \cellcolor[RGB]{242,132,75}\textcolor{black}{.231} & \cellcolor[RGB]{253,196,39}\textcolor{black}{.242} & \cellcolor[RGB]{250,160,57}\textcolor{black}{.236} & \cellcolor[RGB]{253,193,40}\textcolor{black}{.241} & \cellcolor[RGB]{87,1,164}\textcolor{white}{.196} & \cellcolor[RGB]{12,7,134}\textcolor{white}{.187} & \cellcolor[RGB]{242,240,38}\textcolor{black}{.249} & \cellcolor[RGB]{205,73,117}\textcolor{black}{.219} \\
L3 & \cellcolor[RGB]{242,241,38}\textcolor{black}{.239} & \cellcolor[RGB]{251,162,56}\textcolor{black}{.226} & \cellcolor[RGB]{239,248,33}\textcolor{black}{\textbf{.241}} & \cellcolor[RGB]{211,81,111}\textcolor{black}{.209} & \cellcolor[RGB]{251,209,36}\textcolor{black}{.234} & \cellcolor[RGB]{165,31,151}\textcolor{white}{.198} & \cellcolor[RGB]{12,7,134}\textcolor{white}{.173} & \cellcolor[RGB]{253,195,40}\textcolor{black}{.232} & \cellcolor[RGB]{240,128,77}\textcolor{black}{.220} \\
L4 & \cellcolor[RGB]{239,248,33}\textcolor{black}{\textbf{.261}} & \cellcolor[RGB]{237,121,82}\textcolor{black}{.208} & \cellcolor[RGB]{253,187,43}\textcolor{black}{.237} & \cellcolor[RGB]{164,30,152}\textcolor{white}{.159} & \cellcolor[RGB]{240,129,77}\textcolor{black}{.212} & \cellcolor[RGB]{43,5,148}\textcolor{white}{.109} & \cellcolor[RGB]{12,7,134}\textcolor{white}{.101} & \cellcolor[RGB]{219,91,103}\textcolor{black}{.192} & \cellcolor[RGB]{244,137,71}\textcolor{black}{.215} \\
L5 & \cellcolor[RGB]{239,248,33}\textcolor{black}{\textbf{.211}} & \cellcolor[RGB]{207,75,116}\textcolor{black}{.157} & \cellcolor[RGB]{252,170,51}\textcolor{black}{.190} & \cellcolor[RGB]{123,2,168}\textcolor{white}{.127} & \cellcolor[RGB]{195,62,127}\textcolor{white}{.152} & \cellcolor[RGB]{150,18,160}\textcolor{white}{.135} & \cellcolor[RGB]{12,7,134}\textcolor{white}{.099} & \cellcolor[RGB]{167,33,151}\textcolor{white}{.140} & \cellcolor[RGB]{240,128,77}\textcolor{black}{.176} \\
\bottomrule
\end{tabular}
\end{table}

\begin{table}[H]
\centering
\small
\caption{EMOPIA~\cite{emopia} 4-class accuracy (chance .25) $\uparrow$}
\label{tab:app-emopia_4class}
\setlength{\tabcolsep}{4pt}
\begin{tabular}{lccccccccc}
\toprule
Level & \rotatebox{55}{\scriptsize baseline} & \rotatebox{55}{\scriptsize chord-0.5} & \rotatebox{55}{\scriptsize chord-0.1} & \rotatebox{55}{\scriptsize chord-1.0} & \rotatebox{55}{\scriptsize phrase} & \rotatebox{55}{\scriptsize chall-scr1} & \rotatebox{55}{\scriptsize chall-scr1.5} & \rotatebox{55}{\scriptsize lakh1x} & \rotatebox{55}{\scriptsize lakh4x} \\
\midrule
L0 & \cellcolor[RGB]{243,134,73}\textcolor{black}{.476} & \cellcolor[RGB]{12,7,134}\textcolor{white}{.449} & \cellcolor[RGB]{150,18,160}\textcolor{white}{.461} & \cellcolor[RGB]{104,0,167}\textcolor{white}{.457} & \cellcolor[RGB]{206,74,117}\textcolor{black}{.469} & \cellcolor[RGB]{231,110,90}\textcolor{black}{.473} & \cellcolor[RGB]{114,0,168}\textcolor{white}{.458} & \cellcolor[RGB]{239,248,33}\textcolor{black}{\textbf{.487}} & \cellcolor[RGB]{251,208,36}\textcolor{black}{.484} \\
L1 & \cellcolor[RGB]{253,184,44}\textcolor{black}{.486} & \cellcolor[RGB]{170,36,148}\textcolor{white}{.460} & \cellcolor[RGB]{253,198,38}\textcolor{black}{.487} & \cellcolor[RGB]{12,7,134}\textcolor{white}{.440} & \cellcolor[RGB]{194,61,128}\textcolor{white}{.465} & \cellcolor[RGB]{180,45,141}\textcolor{white}{.462} & \cellcolor[RGB]{134,7,166}\textcolor{white}{.455} & \cellcolor[RGB]{239,248,33}\textcolor{black}{\textbf{.494}} & \cellcolor[RGB]{244,234,38}\textcolor{black}{.492} \\
L2 & \cellcolor[RGB]{236,120,83}\textcolor{black}{.483} & \cellcolor[RGB]{191,57,130}\textcolor{white}{.474} & \cellcolor[RGB]{252,170,51}\textcolor{black}{.488} & \cellcolor[RGB]{12,7,134}\textcolor{white}{.457} & \cellcolor[RGB]{214,85,109}\textcolor{black}{.478} & \cellcolor[RGB]{196,63,126}\textcolor{white}{.475} & \cellcolor[RGB]{139,9,164}\textcolor{white}{.468} & \cellcolor[RGB]{239,248,33}\textcolor{black}{\textbf{.496}} & \cellcolor[RGB]{251,162,56}\textcolor{black}{.487} \\
L3 & \cellcolor[RGB]{103,0,167}\textcolor{white}{.467} & \cellcolor[RGB]{253,184,44}\textcolor{black}{.483} & \cellcolor[RGB]{212,82,110}\textcolor{black}{.475} & \cellcolor[RGB]{71,2,159}\textcolor{white}{.465} & \cellcolor[RGB]{250,156,59}\textcolor{black}{.481} & \cellcolor[RGB]{12,7,134}\textcolor{white}{.462} & \cellcolor[RGB]{252,169,52}\textcolor{black}{.482} & \cellcolor[RGB]{239,248,33}\textcolor{black}{\textbf{.486}} & \cellcolor[RGB]{118,1,168}\textcolor{white}{.468} \\
L4 & \cellcolor[RGB]{207,75,116}\textcolor{black}{.444} & \cellcolor[RGB]{236,120,83}\textcolor{black}{.451} & \cellcolor[RGB]{12,7,134}\textcolor{white}{.417} & \cellcolor[RGB]{236,120,83}\textcolor{black}{.451} & \cellcolor[RGB]{239,248,33}\textcolor{black}{\textbf{.468}} & \cellcolor[RGB]{239,126,78}\textcolor{black}{.452} & \cellcolor[RGB]{207,75,116}\textcolor{black}{.444} & \cellcolor[RGB]{253,179,46}\textcolor{black}{.459} & \cellcolor[RGB]{253,187,43}\textcolor{black}{.460} \\
L5 & \cellcolor[RGB]{149,17,161}\textcolor{white}{.403} & \cellcolor[RGB]{239,248,33}\textcolor{black}{\textbf{.426}} & \cellcolor[RGB]{12,7,134}\textcolor{white}{.393} & \cellcolor[RGB]{106,0,167}\textcolor{white}{.400} & \cellcolor[RGB]{192,59,129}\textcolor{white}{.408} & \cellcolor[RGB]{206,74,117}\textcolor{black}{.410} & \cellcolor[RGB]{106,0,167}\textcolor{white}{.400} & \cellcolor[RGB]{247,146,65}\textcolor{black}{.417} & \cellcolor[RGB]{252,199,38}\textcolor{black}{.422} \\
\bottomrule
\end{tabular}
\end{table}

\begin{table}[H]
\centering
\small
\caption{EMOPIA~\cite{emopia} arousal (binary) $\uparrow$}
\label{tab:app-emopia_arousal}
\setlength{\tabcolsep}{4pt}
\begin{tabular}{lccccccccc}
\toprule
Level & \rotatebox{55}{\scriptsize baseline} & \rotatebox{55}{\scriptsize chord-0.5} & \rotatebox{55}{\scriptsize chord-0.1} & \rotatebox{55}{\scriptsize chord-1.0} & \rotatebox{55}{\scriptsize phrase} & \rotatebox{55}{\scriptsize chall-scr1} & \rotatebox{55}{\scriptsize chall-scr1.5} & \rotatebox{55}{\scriptsize lakh1x} & \rotatebox{55}{\scriptsize lakh4x} \\
\midrule
L0 & \cellcolor[RGB]{245,141,69}\textcolor{black}{.765} & \cellcolor[RGB]{71,2,159}\textcolor{white}{.750} & \cellcolor[RGB]{234,116,85}\textcolor{black}{.763} & \cellcolor[RGB]{239,248,33}\textcolor{black}{\textbf{.771}} & \cellcolor[RGB]{234,116,85}\textcolor{black}{.763} & \cellcolor[RGB]{71,2,159}\textcolor{white}{.750} & \cellcolor[RGB]{12,7,134}\textcolor{white}{.747} & \cellcolor[RGB]{172,37,147}\textcolor{white}{.756} & \cellcolor[RGB]{159,26,155}\textcolor{white}{.755} \\
L1 & \cellcolor[RGB]{246,143,67}\textcolor{black}{.765} & \cellcolor[RGB]{217,89,105}\textcolor{black}{.761} & \cellcolor[RGB]{251,211,36}\textcolor{black}{.768} & \cellcolor[RGB]{197,64,125}\textcolor{white}{.759} & \cellcolor[RGB]{239,248,33}\textcolor{black}{\textbf{.770}} & \cellcolor[RGB]{95,0,166}\textcolor{white}{.753} & \cellcolor[RGB]{12,7,134}\textcolor{white}{.749} & \cellcolor[RGB]{246,229,37}\textcolor{black}{.769} & \cellcolor[RGB]{197,64,125}\textcolor{white}{.759} \\
L2 & \cellcolor[RGB]{250,160,57}\textcolor{black}{.769} & \cellcolor[RGB]{163,29,153}\textcolor{white}{.764} & \cellcolor[RGB]{243,135,72}\textcolor{black}{.768} & \cellcolor[RGB]{163,29,153}\textcolor{white}{.764} & \cellcolor[RGB]{82,1,163}\textcolor{white}{.761} & \cellcolor[RGB]{12,7,134}\textcolor{white}{.759} & \cellcolor[RGB]{137,8,165}\textcolor{white}{.763} & \cellcolor[RGB]{239,248,33}\textcolor{black}{\textbf{.772}} & \cellcolor[RGB]{232,113,88}\textcolor{black}{.768} \\
L3 & \cellcolor[RGB]{236,120,83}\textcolor{black}{.762} & \cellcolor[RGB]{178,44,142}\textcolor{white}{.755} & \cellcolor[RGB]{252,172,50}\textcolor{black}{.766} & \cellcolor[RGB]{207,75,116}\textcolor{black}{.758} & \cellcolor[RGB]{239,248,33}\textcolor{black}{\textbf{.770}} & \cellcolor[RGB]{142,12,164}\textcolor{white}{.753} & \cellcolor[RGB]{12,7,134}\textcolor{white}{.745} & \cellcolor[RGB]{252,201,38}\textcolor{black}{.768} & \cellcolor[RGB]{253,185,43}\textcolor{black}{.767} \\
L4 & \cellcolor[RGB]{252,206,37}\textcolor{black}{.767} & \cellcolor[RGB]{249,154,60}\textcolor{black}{.763} & \cellcolor[RGB]{51,4,151}\textcolor{white}{.743} & \cellcolor[RGB]{167,33,151}\textcolor{white}{.752} & \cellcolor[RGB]{239,248,33}\textcolor{black}{\textbf{.769}} & \cellcolor[RGB]{176,42,143}\textcolor{white}{.753} & \cellcolor[RGB]{12,7,134}\textcolor{white}{.741} & \cellcolor[RGB]{218,90,104}\textcolor{black}{.757} & \cellcolor[RGB]{252,206,37}\textcolor{black}{.767} \\
L5 & \cellcolor[RGB]{251,163,55}\textcolor{black}{.758} & \cellcolor[RGB]{246,142,68}\textcolor{black}{.754} & \cellcolor[RGB]{253,187,43}\textcolor{black}{.763} & \cellcolor[RGB]{165,31,151}\textcolor{white}{.725} & \cellcolor[RGB]{248,150,63}\textcolor{black}{.755} & \cellcolor[RGB]{173,38,146}\textcolor{white}{.727} & \cellcolor[RGB]{12,7,134}\textcolor{white}{.697} & \cellcolor[RGB]{239,248,33}\textcolor{black}{\textbf{.774}} & \cellcolor[RGB]{248,150,63}\textcolor{black}{.755} \\
\bottomrule
\end{tabular}
\end{table}

\begin{table}[H]
\centering
\small
\caption{EMOPIA~\cite{emopia} valence (binary) $\uparrow$}
\label{tab:app-emopia_valence}
\setlength{\tabcolsep}{4pt}
\begin{tabular}{lccccccccc}
\toprule
Level & \rotatebox{55}{\scriptsize baseline} & \rotatebox{55}{\scriptsize chord-0.5} & \rotatebox{55}{\scriptsize chord-0.1} & \rotatebox{55}{\scriptsize chord-1.0} & \rotatebox{55}{\scriptsize phrase} & \rotatebox{55}{\scriptsize chall-scr1} & \rotatebox{55}{\scriptsize chall-scr1.5} & \rotatebox{55}{\scriptsize lakh1x} & \rotatebox{55}{\scriptsize lakh4x} \\
\midrule
L0 & \cellcolor[RGB]{188,54,133}\textcolor{white}{.600} & \cellcolor[RGB]{215,87,107}\textcolor{black}{.602} & \cellcolor[RGB]{12,7,134}\textcolor{white}{.593} & \cellcolor[RGB]{155,23,158}\textcolor{white}{.599} & \cellcolor[RGB]{189,55,132}\textcolor{white}{.600} & \cellcolor[RGB]{45,4,148}\textcolor{white}{.594} & \cellcolor[RGB]{245,139,70}\textcolor{black}{.605} & \cellcolor[RGB]{155,23,158}\textcolor{white}{.599} & \cellcolor[RGB]{239,248,33}\textcolor{black}{\textbf{.610}} \\
L1 & \cellcolor[RGB]{159,26,155}\textcolor{white}{.611} & \cellcolor[RGB]{239,248,33}\textcolor{black}{\textbf{.639}} & \cellcolor[RGB]{121,1,168}\textcolor{white}{.606} & \cellcolor[RGB]{12,7,134}\textcolor{white}{.596} & \cellcolor[RGB]{191,57,130}\textcolor{white}{.615} & \cellcolor[RGB]{104,0,167}\textcolor{white}{.604} & \cellcolor[RGB]{160,27,155}\textcolor{white}{.611} & \cellcolor[RGB]{112,0,168}\textcolor{white}{.605} & \cellcolor[RGB]{213,83,109}\textcolor{black}{.619} \\
L2 & \cellcolor[RGB]{253,187,43}\textcolor{black}{.627} & \cellcolor[RGB]{251,211,36}\textcolor{black}{.628} & \cellcolor[RGB]{251,211,36}\textcolor{black}{.628} & \cellcolor[RGB]{97,0,166}\textcolor{white}{.605} & \cellcolor[RGB]{162,28,154}\textcolor{white}{.611} & \cellcolor[RGB]{253,187,43}\textcolor{black}{.626} & \cellcolor[RGB]{151,19,160}\textcolor{white}{.610} & \cellcolor[RGB]{239,248,33}\textcolor{black}{\textbf{.631}} & \cellcolor[RGB]{12,7,134}\textcolor{white}{.599} \\
L3 & \cellcolor[RGB]{249,154,60}\textcolor{black}{.618} & \cellcolor[RGB]{249,153,61}\textcolor{black}{.618} & \cellcolor[RGB]{243,236,38}\textcolor{black}{.625} & \cellcolor[RGB]{253,176,48}\textcolor{black}{.620} & \cellcolor[RGB]{239,248,33}\textcolor{black}{\textbf{.626}} & \cellcolor[RGB]{12,7,134}\textcolor{white}{.594} & \cellcolor[RGB]{228,106,93}\textcolor{black}{.613} & \cellcolor[RGB]{252,199,38}\textcolor{black}{.622} & \cellcolor[RGB]{130,4,167}\textcolor{white}{.602} \\
L4 & \cellcolor[RGB]{217,89,105}\textcolor{black}{.585} & \cellcolor[RGB]{12,7,134}\textcolor{white}{.552} & \cellcolor[RGB]{74,2,160}\textcolor{white}{.559} & \cellcolor[RGB]{207,75,116}\textcolor{black}{.583} & \cellcolor[RGB]{239,248,33}\textcolor{black}{\textbf{.612}} & \cellcolor[RGB]{182,47,139}\textcolor{white}{.577} & \cellcolor[RGB]{154,21,158}\textcolor{white}{.571} & \cellcolor[RGB]{242,132,75}\textcolor{black}{.594} & \cellcolor[RGB]{250,159,58}\textcolor{black}{.598} \\
L5 & \cellcolor[RGB]{12,7,134}\textcolor{white}{.535} & \cellcolor[RGB]{251,163,55}\textcolor{black}{.553} & \cellcolor[RGB]{58,4,154}\textcolor{white}{.537} & \cellcolor[RGB]{212,82,110}\textcolor{black}{.547} & \cellcolor[RGB]{192,59,129}\textcolor{white}{.545} & \cellcolor[RGB]{221,95,101}\textcolor{black}{.548} & \cellcolor[RGB]{243,134,73}\textcolor{black}{.551} & \cellcolor[RGB]{239,248,33}\textcolor{black}{\textbf{.557}} & \cellcolor[RGB]{253,195,40}\textcolor{black}{.555} \\
\bottomrule
\end{tabular}
\end{table}

\section{Equivariance}
\label{app:equiv}
$R^2$ of a linear fit of per-level embedding distance versus shift magnitude,
computed from the probes' logged distance curves (higher = distances more
predictably ordered by shift size).

\begin{table}[H]
\centering
\small
\caption{Pitch-transposition equivariance $R^2$ $\uparrow$}
\label{tab:app-pitch_transposition}
\setlength{\tabcolsep}{4pt}
\begin{tabular}{lccccccccc}
\toprule
Level & \rotatebox{55}{\scriptsize baseline} & \rotatebox{55}{\scriptsize chord-0.5} & \rotatebox{55}{\scriptsize chord-0.1} & \rotatebox{55}{\scriptsize chord-1.0} & \rotatebox{55}{\scriptsize phrase} & \rotatebox{55}{\scriptsize chall-scr1} & \rotatebox{55}{\scriptsize chall-scr1.5} & \rotatebox{55}{\scriptsize lakh1x} & \rotatebox{55}{\scriptsize lakh4x} \\
\midrule
L0 & \cellcolor[RGB]{239,248,33}\textcolor{black}{\textbf{1.000}} & \cellcolor[RGB]{252,172,50}\textcolor{black}{.999} & \cellcolor[RGB]{139,9,164}\textcolor{white}{.996} & \cellcolor[RGB]{12,7,134}\textcolor{white}{.995} & \cellcolor[RGB]{253,188,42}\textcolor{black}{.999} & \cellcolor[RGB]{248,150,63}\textcolor{black}{.999} & \cellcolor[RGB]{203,71,119}\textcolor{black}{.997} & \cellcolor[RGB]{232,112,89}\textcolor{black}{.998} & \cellcolor[RGB]{253,198,38}\textcolor{black}{.999} \\
L1 & \cellcolor[RGB]{251,209,36}\textcolor{black}{.997} & \cellcolor[RGB]{248,152,62}\textcolor{black}{.995} & \cellcolor[RGB]{193,60,128}\textcolor{white}{.991} & \cellcolor[RGB]{12,7,134}\textcolor{white}{.986} & \cellcolor[RGB]{252,166,53}\textcolor{black}{.996} & \cellcolor[RGB]{251,164,54}\textcolor{black}{.996} & \cellcolor[RGB]{228,106,93}\textcolor{black}{.993} & \cellcolor[RGB]{229,108,91}\textcolor{black}{.994} & \cellcolor[RGB]{239,248,33}\textcolor{black}{\textbf{.998}} \\
L2 & \cellcolor[RGB]{223,97,99}\textcolor{black}{.991} & \cellcolor[RGB]{227,103,95}\textcolor{black}{.992} & \cellcolor[RGB]{109,0,168}\textcolor{white}{.986} & \cellcolor[RGB]{45,4,148}\textcolor{white}{.984} & \cellcolor[RGB]{208,77,115}\textcolor{black}{.990} & \cellcolor[RGB]{143,13,163}\textcolor{white}{.987} & \cellcolor[RGB]{86,1,163}\textcolor{white}{.985} & \cellcolor[RGB]{12,7,134}\textcolor{white}{.983} & \cellcolor[RGB]{239,248,33}\textcolor{black}{\textbf{.997}} \\
L3 & \cellcolor[RGB]{242,241,38}\textcolor{black}{.990} & \cellcolor[RGB]{246,145,66}\textcolor{black}{.985} & \cellcolor[RGB]{253,190,41}\textcolor{black}{.988} & \cellcolor[RGB]{86,1,163}\textcolor{white}{.974} & \cellcolor[RGB]{239,248,33}\textcolor{black}{\textbf{.991}} & \cellcolor[RGB]{201,69,121}\textcolor{black}{.981} & \cellcolor[RGB]{12,7,134}\textcolor{white}{.971} & \cellcolor[RGB]{250,214,36}\textcolor{black}{.989} & \cellcolor[RGB]{243,236,38}\textcolor{black}{.990} \\
L4 & \cellcolor[RGB]{240,246,35}\textcolor{black}{.963} & \cellcolor[RGB]{252,199,38}\textcolor{black}{.871} & \cellcolor[RGB]{240,246,35}\textcolor{black}{.963} & \cellcolor[RGB]{252,169,52}\textcolor{black}{.805} & \cellcolor[RGB]{239,248,33}\textcolor{black}{.966} & \cellcolor[RGB]{12,7,134}\textcolor{white}{.125} & \cellcolor[RGB]{59,3,154}\textcolor{white}{.199} & \cellcolor[RGB]{240,246,35}\textcolor{black}{.963} & \cellcolor[RGB]{239,248,33}\textcolor{black}{\textbf{.968}} \\
L5 & \cellcolor[RGB]{248,221,36}\textcolor{black}{.878} & \cellcolor[RGB]{239,125,79}\textcolor{black}{.747} & \cellcolor[RGB]{250,216,36}\textcolor{black}{.872} & \cellcolor[RGB]{223,98,98}\textcolor{black}{.700} & \cellcolor[RGB]{239,248,33}\textcolor{black}{\textbf{.912}} & \cellcolor[RGB]{94,0,165}\textcolor{white}{.478} & \cellcolor[RGB]{12,7,134}\textcolor{white}{.388} & \cellcolor[RGB]{253,185,43}\textcolor{black}{.834} & \cellcolor[RGB]{245,233,38}\textcolor{black}{.893} \\
\bottomrule
\end{tabular}
\end{table}

\begin{table}[H]
\centering
\small
\caption{Time-translation equivariance $R^2$ $\uparrow$}
\label{tab:app-time_translation}
\setlength{\tabcolsep}{4pt}
\begin{tabular}{lccccccccc}
\toprule
Level & \rotatebox{55}{\scriptsize baseline} & \rotatebox{55}{\scriptsize chord-0.5} & \rotatebox{55}{\scriptsize chord-0.1} & \rotatebox{55}{\scriptsize chord-1.0} & \rotatebox{55}{\scriptsize phrase} & \rotatebox{55}{\scriptsize chall-scr1} & \rotatebox{55}{\scriptsize chall-scr1.5} & \rotatebox{55}{\scriptsize lakh1x} & \rotatebox{55}{\scriptsize lakh4x} \\
\midrule
L0 & \cellcolor[RGB]{245,231,38}\textcolor{black}{.980} & \cellcolor[RGB]{115,0,168}\textcolor{white}{.938} & \cellcolor[RGB]{240,128,77}\textcolor{black}{.964} & \cellcolor[RGB]{76,2,161}\textcolor{white}{.933} & \cellcolor[RGB]{249,219,36}\textcolor{black}{.978} & \cellcolor[RGB]{12,7,134}\textcolor{white}{.925} & \cellcolor[RGB]{168,34,150}\textcolor{white}{.946} & \cellcolor[RGB]{239,248,33}\textcolor{black}{\textbf{.982}} & \cellcolor[RGB]{240,246,35}\textcolor{black}{.981} \\
L1 & \cellcolor[RGB]{249,154,60}\textcolor{black}{.911} & \cellcolor[RGB]{130,4,167}\textcolor{white}{.801} & \cellcolor[RGB]{246,145,66}\textcolor{black}{.905} & \cellcolor[RGB]{120,1,168}\textcolor{white}{.795} & \cellcolor[RGB]{252,170,51}\textcolor{black}{.920} & \cellcolor[RGB]{64,3,156}\textcolor{white}{.765} & \cellcolor[RGB]{12,7,134}\textcolor{white}{.743} & \cellcolor[RGB]{252,206,37}\textcolor{black}{.940} & \cellcolor[RGB]{239,248,33}\textcolor{black}{\textbf{.962}} \\
L2 & \cellcolor[RGB]{246,145,66}\textcolor{black}{.848} & \cellcolor[RGB]{130,4,167}\textcolor{white}{.680} & \cellcolor[RGB]{242,132,75}\textcolor{black}{.835} & \cellcolor[RGB]{160,27,155}\textcolor{white}{.710} & \cellcolor[RGB]{248,149,64}\textcolor{black}{.853} & \cellcolor[RGB]{92,0,165}\textcolor{white}{.644} & \cellcolor[RGB]{12,7,134}\textcolor{white}{.586} & \cellcolor[RGB]{252,204,37}\textcolor{black}{.905} & \cellcolor[RGB]{239,248,33}\textcolor{black}{\textbf{.941}} \\
L3 & \cellcolor[RGB]{251,164,54}\textcolor{black}{.730} & \cellcolor[RGB]{124,2,167}\textcolor{white}{.569} & \cellcolor[RGB]{251,208,36}\textcolor{black}{.762} & \cellcolor[RGB]{120,1,168}\textcolor{white}{.565} & \cellcolor[RGB]{253,193,40}\textcolor{black}{.751} & \cellcolor[RGB]{127,3,167}\textcolor{white}{.571} & \cellcolor[RGB]{12,7,134}\textcolor{white}{.495} & \cellcolor[RGB]{243,236,38}\textcolor{black}{.782} & \cellcolor[RGB]{239,248,33}\textcolor{black}{\textbf{.790}} \\
L4 & \cellcolor[RGB]{251,209,36}\textcolor{black}{.539} & \cellcolor[RGB]{196,63,126}\textcolor{white}{.431} & \cellcolor[RGB]{252,203,37}\textcolor{black}{.536} & \cellcolor[RGB]{157,24,157}\textcolor{white}{.398} & \cellcolor[RGB]{253,195,40}\textcolor{black}{.530} & \cellcolor[RGB]{12,7,134}\textcolor{white}{.314} & \cellcolor[RGB]{39,5,146}\textcolor{white}{.325} & \cellcolor[RGB]{250,156,59}\textcolor{black}{.506} & \cellcolor[RGB]{239,248,33}\textcolor{black}{\textbf{.562}} \\
L5 & \cellcolor[RGB]{253,185,43}\textcolor{black}{.408} & \cellcolor[RGB]{188,54,133}\textcolor{white}{.366} & \cellcolor[RGB]{252,169,52}\textcolor{black}{.403} & \cellcolor[RGB]{126,3,167}\textcolor{white}{.347} & \cellcolor[RGB]{239,248,33}\textcolor{black}{\textbf{.423}} & \cellcolor[RGB]{81,1,162}\textcolor{white}{.335} & \cellcolor[RGB]{12,7,134}\textcolor{white}{.321} & \cellcolor[RGB]{235,118,84}\textcolor{black}{.388} & \cellcolor[RGB]{252,201,38}\textcolor{black}{.412} \\
\bottomrule
\end{tabular}
\end{table}

\begin{figure}[tb]
\centering
\includegraphics[width=0.85\columnwidth]{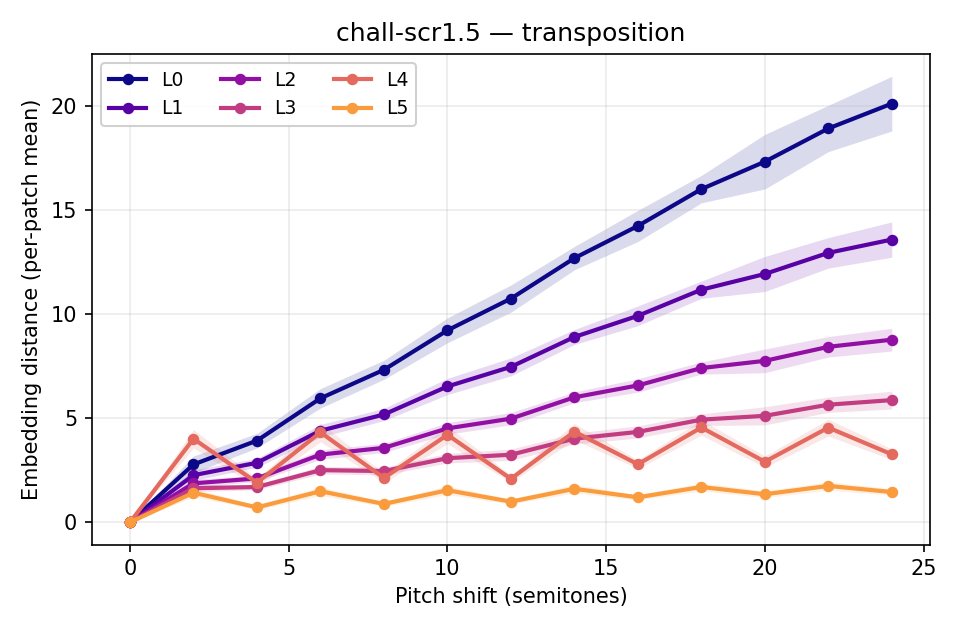}
\caption{Per-level embedding distance versus pitch transposition shift for \texttt{chall-scr1.5}; shaded bands are $\pm$1 std over probe samples.}
\label{fig:app-transposition-chall-scr1.5}
\end{figure}

\begin{figure}[tb]
\centering
\includegraphics[width=0.85\columnwidth]{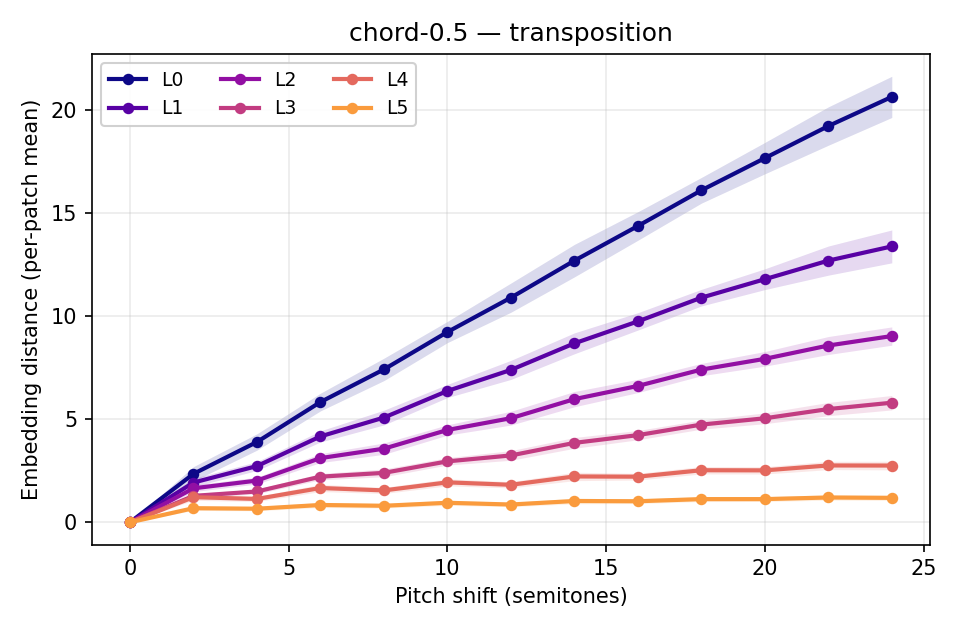}
\caption{Per-level embedding distance versus pitch transposition shift for \texttt{chord-0.5}; shaded bands are $\pm$1 std over probe samples.}
\label{fig:app-transposition-chord-0.5}
\end{figure}

\begin{figure}[tb]
\centering
\includegraphics[width=0.85\columnwidth]{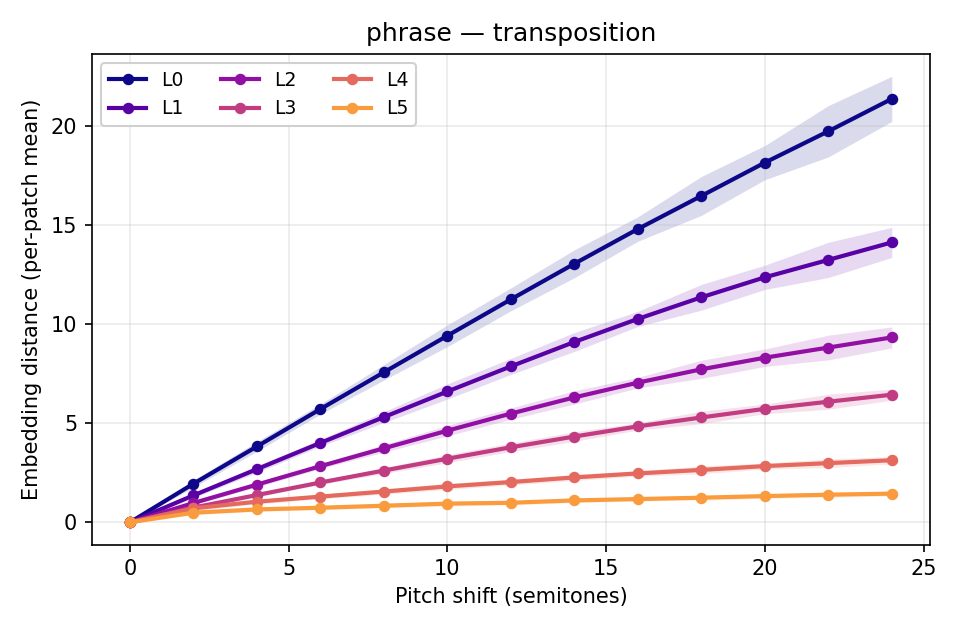}
\caption{Per-level embedding distance versus pitch transposition shift for \texttt{phrase}; shaded bands are $\pm$1 std over probe samples.}
\label{fig:app-transposition-phrase}
\end{figure}

\begin{figure}[tb]
\centering
\includegraphics[width=0.85\columnwidth]{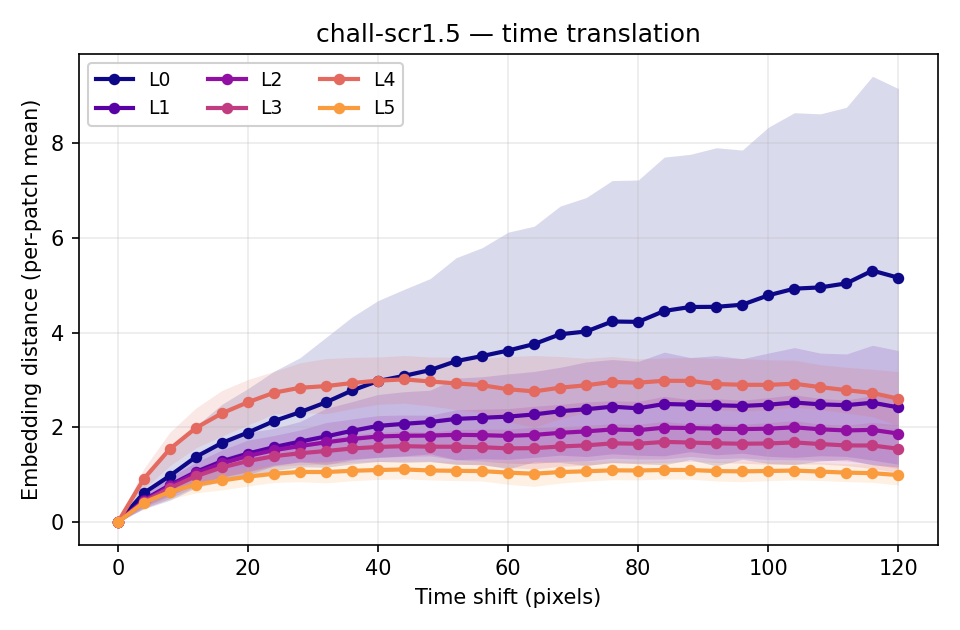}
\caption{Per-level embedding distance versus time translation shift for \texttt{chall-scr1.5}; shaded bands are $\pm$1 std over probe samples.}
\label{fig:app-time_translation-chall-scr1.5}
\end{figure}

\begin{figure}[tb]
\centering
\includegraphics[width=0.85\columnwidth]{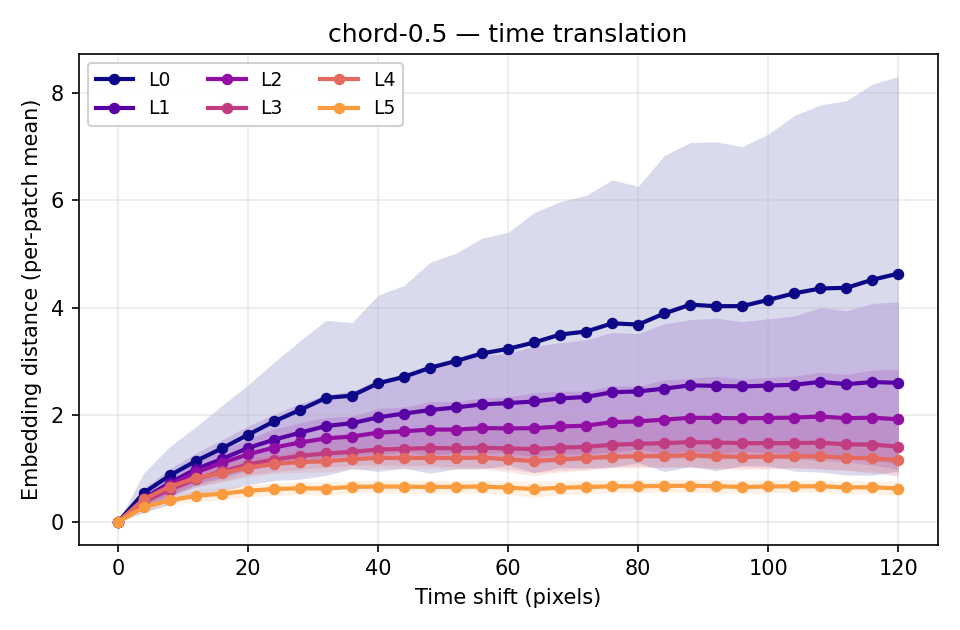}
\caption{Per-level embedding distance versus time translation shift for \texttt{chord-0.5}; shaded bands are $\pm$1 std over probe samples.}
\label{fig:app-time_translation-chord-0.5}
\end{figure}

\begin{figure}[tb]
\centering
\includegraphics[width=0.85\columnwidth]{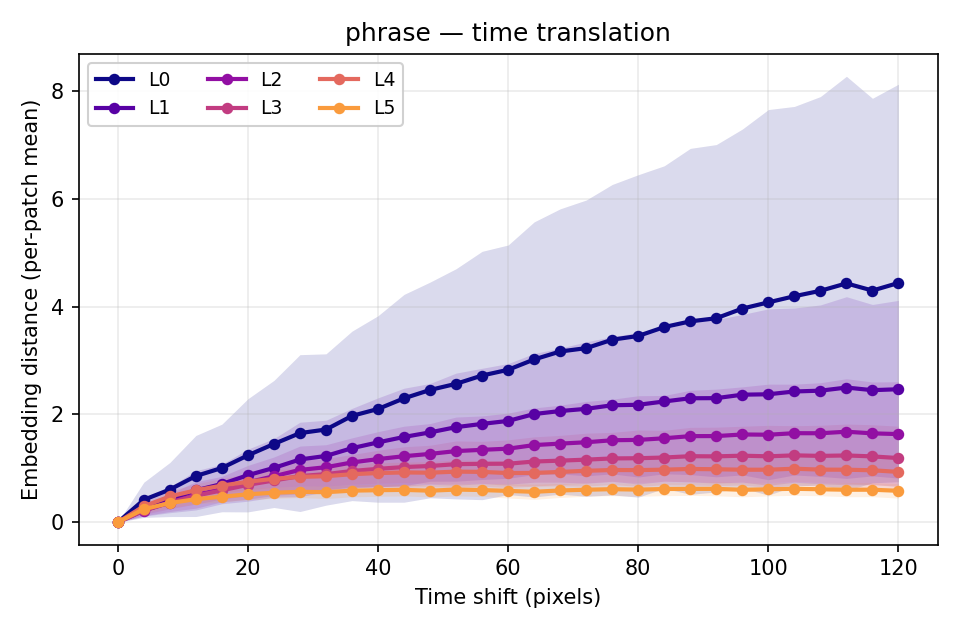}
\caption{Per-level embedding distance versus time translation shift for \texttt{phrase}; shaded bands are $\pm$1 std over probe samples.}
\label{fig:app-time_translation-phrase}
\end{figure}

\section{Phrase-boundary analyses}
\label{app:phrase}
Two measurements against the human phrase annotations of Dai et al.: (1) the
unsupervised Foote-novelty segmentation F1 (tolerance in columns of 32nd-note
resolution; chance F1@32 $\approx$ .275 for count-matched random placement,
fixed 8-bar grid .405, human inter-annotator ceiling .895), and (2) the linear
boundary-detection probe from the main text, per level (AP chance = .17 crop
base rate; AUC chance = .50; raw 32$\times$32 pixels score .18/.51).

\begin{table}[H]
\centering
\small
\caption{Novelty-pipeline boundary F1 by encoder, tolerance, and level.}
\label{tab:app-phrase-f1}
\setlength{\tabcolsep}{4pt}
\begin{tabular}{lccccccc}
\toprule
 & L0 & L1 & L2 & L3 & L4 & L5 & pixel \\
\midrule
baseline F1@16 & \cellcolor[RGB]{239,248,33}\textcolor{black}{\textbf{.170}} & \cellcolor[RGB]{253,176,48}\textcolor{black}{.149} & \cellcolor[RGB]{253,181,45}\textcolor{black}{.150} & \cellcolor[RGB]{252,173,49}\textcolor{black}{.148} & \cellcolor[RGB]{230,109,90}\textcolor{black}{.125} & \cellcolor[RGB]{184,50,137}\textcolor{white}{.100} & \cellcolor[RGB]{12,7,134}\textcolor{white}{.048} \\
baseline F1@32 & \cellcolor[RGB]{239,248,33}\textcolor{black}{\textbf{.303}} & \cellcolor[RGB]{253,195,40}\textcolor{black}{.277} & \cellcolor[RGB]{246,142,68}\textcolor{black}{.248} & \cellcolor[RGB]{232,112,89}\textcolor{black}{.229} & \cellcolor[RGB]{203,71,119}\textcolor{black}{.201} & \cellcolor[RGB]{178,44,142}\textcolor{white}{.182} & \cellcolor[RGB]{12,7,134}\textcolor{white}{.098} \\
chord-0.5 F1@16 & \cellcolor[RGB]{239,248,33}\textcolor{black}{\textbf{.134}} & \cellcolor[RGB]{250,159,58}\textcolor{black}{.115} & \cellcolor[RGB]{149,17,161}\textcolor{white}{.075} & \cellcolor[RGB]{253,181,45}\textcolor{black}{.120} & \cellcolor[RGB]{249,154,60}\textcolor{black}{.114} & \cellcolor[RGB]{208,77,115}\textcolor{black}{.093} & \cellcolor[RGB]{12,7,134}\textcolor{white}{.048} \\
chord-0.5 F1@32 & \cellcolor[RGB]{239,248,33}\textcolor{black}{\textbf{.246}} & \cellcolor[RGB]{253,188,42}\textcolor{black}{.225} & \cellcolor[RGB]{187,53,134}\textcolor{white}{.163} & \cellcolor[RGB]{242,132,75}\textcolor{black}{.202} & \cellcolor[RGB]{229,108,91}\textcolor{black}{.191} & \cellcolor[RGB]{190,56,131}\textcolor{white}{.165} & \cellcolor[RGB]{12,7,134}\textcolor{white}{.098} \\
phrase-encoder F1@16 & \cellcolor[RGB]{239,248,33}\textcolor{black}{\textbf{.179}} & \cellcolor[RGB]{252,170,51}\textcolor{black}{.154} & \cellcolor[RGB]{253,184,44}\textcolor{black}{.159} & \cellcolor[RGB]{243,134,73}\textcolor{black}{.141} & \cellcolor[RGB]{219,91,103}\textcolor{black}{.123} & \cellcolor[RGB]{175,40,144}\textcolor{white}{.100} & \cellcolor[RGB]{12,7,134}\textcolor{white}{.048} \\
phrase-encoder F1@32 & \cellcolor[RGB]{239,248,33}\textcolor{black}{\textbf{.308}} & \cellcolor[RGB]{253,181,45}\textcolor{black}{.274} & \cellcolor[RGB]{240,128,77}\textcolor{black}{.243} & \cellcolor[RGB]{229,108,91}\textcolor{black}{.230} & \cellcolor[RGB]{198,65,124}\textcolor{white}{.199} & \cellcolor[RGB]{192,59,129}\textcolor{white}{.194} & \cellcolor[RGB]{12,7,134}\textcolor{white}{.098} \\
\bottomrule
\end{tabular}
\end{table}

\begin{table}[H]
\centering
\small
\caption{Linear boundary-detection probe, AP (chance = .17).}
\label{tab:app-bprobe-ap}
\setlength{\tabcolsep}{4pt}
\begin{tabular}{lccccccc}
\toprule
 & L0 & L1 & L2 & L3 & L4 & L5 & pixels \\
\midrule
MRJS & \cellcolor[RGB]{239,248,33}\textcolor{black}{\textbf{.273}} & \cellcolor[RGB]{245,233,38}\textcolor{black}{.269} & \cellcolor[RGB]{252,204,37}\textcolor{black}{.263} & \cellcolor[RGB]{149,17,161}\textcolor{white}{.210} & \cellcolor[RGB]{129,4,167}\textcolor{white}{.205} & \cellcolor[RGB]{66,3,157}\textcolor{white}{.191} & \cellcolor[RGB]{12,7,134}\textcolor{white}{.181} \\
+chords & \cellcolor[RGB]{251,209,36}\textcolor{black}{.241} & \cellcolor[RGB]{253,195,40}\textcolor{black}{.239} & \cellcolor[RGB]{239,248,33}\textcolor{black}{\textbf{.248}} & \cellcolor[RGB]{130,4,167}\textcolor{white}{.194} & \cellcolor[RGB]{76,2,161}\textcolor{white}{.184} & \cellcolor[RGB]{12,7,134}\textcolor{white}{.175} & \cellcolor[RGB]{58,4,154}\textcolor{white}{.181} \\
+phrases & \cellcolor[RGB]{253,195,40}\textcolor{black}{.279} & \cellcolor[RGB]{252,203,37}\textcolor{black}{.282} & \cellcolor[RGB]{239,248,33}\textcolor{black}{\textbf{.294}} & \cellcolor[RGB]{177,43,143}\textcolor{white}{.224} & \cellcolor[RGB]{12,7,134}\textcolor{white}{.177} & \cellcolor[RGB]{58,4,154}\textcolor{white}{.186} & \cellcolor[RGB]{35,5,144}\textcolor{white}{.181} \\
Lakh1x & \cellcolor[RGB]{253,185,43}\textcolor{black}{.267} & \cellcolor[RGB]{239,248,33}\textcolor{black}{\textbf{.283}} & \cellcolor[RGB]{252,206,37}\textcolor{black}{.272} & \cellcolor[RGB]{249,153,61}\textcolor{black}{.257} & \cellcolor[RGB]{12,7,134}\textcolor{white}{.171} & \cellcolor[RGB]{90,0,165}\textcolor{white}{.189} & \cellcolor[RGB]{61,3,155}\textcolor{white}{.181} \\
Lakh4x & \cellcolor[RGB]{239,248,33}\textcolor{black}{\textbf{.262}} & \cellcolor[RGB]{240,245,37}\textcolor{black}{.261} & \cellcolor[RGB]{246,228,37}\textcolor{black}{.258} & \cellcolor[RGB]{170,36,148}\textcolor{white}{.212} & \cellcolor[RGB]{63,3,156}\textcolor{white}{.189} & \cellcolor[RGB]{35,5,144}\textcolor{white}{.184} & \cellcolor[RGB]{12,7,134}\textcolor{white}{.181} \\
MRJ48 & \cellcolor[RGB]{190,56,131}\textcolor{white}{.224} & \cellcolor[RGB]{239,248,33}\textcolor{black}{\textbf{.277}} & \cellcolor[RGB]{252,203,37}\textcolor{black}{.267} & \cellcolor[RGB]{183,48,138}\textcolor{white}{.222} & \cellcolor[RGB]{16,7,135}\textcolor{white}{.182} & \cellcolor[RGB]{90,0,165}\textcolor{white}{.196} & \cellcolor[RGB]{12,7,134}\textcolor{white}{.181} \\
DINOv2 & \cellcolor[RGB]{239,248,33}\textcolor{black}{\textbf{.210}} & \cellcolor[RGB]{236,120,83}\textcolor{black}{.200} & --- & --- & --- & --- & \cellcolor[RGB]{12,7,134}\textcolor{white}{.181} \\
\bottomrule
\end{tabular}
\end{table}

\begin{table}[H]
\centering
\small
\caption{Linear boundary-detection probe, AUC (chance = .50).}
\label{tab:app-bprobe-auc}
\setlength{\tabcolsep}{4pt}
\begin{tabular}{lccccccc}
\toprule
 & L0 & L1 & L2 & L3 & L4 & L5 & pixels \\
\midrule
MRJS & \cellcolor[RGB]{239,248,33}\textcolor{black}{\textbf{.613}} & \cellcolor[RGB]{252,206,37}\textcolor{black}{.603} & \cellcolor[RGB]{252,173,49}\textcolor{black}{.594} & \cellcolor[RGB]{207,75,116}\textcolor{black}{.562} & \cellcolor[RGB]{158,25,156}\textcolor{white}{.544} & \cellcolor[RGB]{45,4,148}\textcolor{white}{.514} & \cellcolor[RGB]{12,7,134}\textcolor{white}{.508} \\
+chords & \cellcolor[RGB]{239,248,33}\textcolor{black}{\textbf{.580}} & \cellcolor[RGB]{250,159,58}\textcolor{black}{.563} & \cellcolor[RGB]{250,213,36}\textcolor{black}{.574} & \cellcolor[RGB]{190,56,131}\textcolor{white}{.537} & \cellcolor[RGB]{66,3,157}\textcolor{white}{.510} & \cellcolor[RGB]{12,7,134}\textcolor{white}{.502} & \cellcolor[RGB]{56,4,153}\textcolor{white}{.508} \\
+phrases & \cellcolor[RGB]{239,248,33}\textcolor{black}{\textbf{.617}} & \cellcolor[RGB]{252,203,37}\textcolor{black}{.605} & \cellcolor[RGB]{252,204,37}\textcolor{black}{.605} & \cellcolor[RGB]{212,82,110}\textcolor{black}{.567} & \cellcolor[RGB]{41,5,147}\textcolor{white}{.513} & \cellcolor[RGB]{39,5,146}\textcolor{white}{.513} & \cellcolor[RGB]{12,7,134}\textcolor{white}{.508} \\
Lakh1x & \cellcolor[RGB]{252,199,38}\textcolor{black}{.594} & \cellcolor[RGB]{239,248,33}\textcolor{black}{\textbf{.608}} & \cellcolor[RGB]{252,169,52}\textcolor{black}{.586} & \cellcolor[RGB]{251,163,55}\textcolor{black}{.583} & \cellcolor[RGB]{12,7,134}\textcolor{white}{.493} & \cellcolor[RGB]{94,0,165}\textcolor{white}{.512} & \cellcolor[RGB]{78,2,161}\textcolor{white}{.508} \\
Lakh4x & \cellcolor[RGB]{239,248,33}\textcolor{black}{\textbf{.614}} & \cellcolor[RGB]{253,178,47}\textcolor{black}{.596} & \cellcolor[RGB]{253,187,43}\textcolor{black}{.598} & \cellcolor[RGB]{201,69,121}\textcolor{black}{.559} & \cellcolor[RGB]{100,0,167}\textcolor{white}{.525} & \cellcolor[RGB]{12,7,134}\textcolor{white}{.505} & \cellcolor[RGB]{31,5,142}\textcolor{white}{.508} \\
MRJ48 & \cellcolor[RGB]{208,77,115}\textcolor{black}{.563} & \cellcolor[RGB]{253,190,41}\textcolor{black}{.598} & \cellcolor[RGB]{239,248,33}\textcolor{black}{\textbf{.613}} & \cellcolor[RGB]{228,106,93}\textcolor{black}{.573} & \cellcolor[RGB]{56,4,153}\textcolor{white}{.517} & \cellcolor[RGB]{106,0,167}\textcolor{white}{.529} & \cellcolor[RGB]{12,7,134}\textcolor{white}{.508} \\
DINOv2 & \cellcolor[RGB]{239,248,33}\textcolor{black}{\textbf{.572}} & \cellcolor[RGB]{233,114,87}\textcolor{black}{.549} & --- & --- & --- & --- & \cellcolor[RGB]{12,7,134}\textcolor{white}{.508} \\
\bottomrule
\end{tabular}
\end{table}

\section{Ablation studies}
\label{app:ablations}
Screening campaigns behind the production recipe. Cells are shaded per
\emph{column} here (variants are rows); $\times$ marks catastrophic L5 chroma
collapse ($R^2$ of $-77$ to $-713$ despite normal training curves), --- marks
missing data. Validation losses are comparable only within a table.

\begin{table}[H]
\centering
\small
\caption{Depth allocation across Swin levels (100-epoch screens; content probes at L5, cross-song = min across levels). $\dagger$ = variant with L5 chroma collapse.}
\label{tab:app-depth}
\setlength{\tabcolsep}{4pt}
\begin{tabular}{lccccccc}
\toprule
Variant & \rotatebox{55}{\scriptsize Blocks} & \rotatebox{55}{\scriptsize Chroma R² L5 ↑} & \rotatebox{55}{\scriptsize Key L5 ↑} & \rotatebox{55}{\scriptsize Root L5 ↑} & \rotatebox{55}{\scriptsize Density L5 ↑} & \rotatebox{55}{\scriptsize Cross-song min ↓} & \rotatebox{55}{\scriptsize Val loss ↓} \\
\midrule
baseline {\tiny [2,2,2,6,2,2]} & \cellcolor[RGB]{239,248,33}\textcolor{black}{\textbf{16}} & \cellcolor[RGB]{248,223,36}\textcolor{black}{.701} & \cellcolor[RGB]{12,7,134}\textcolor{white}{.178} & \cellcolor[RGB]{144,14,163}\textcolor{white}{.264} & \cellcolor[RGB]{35,5,144}\textcolor{white}{.893} & \cellcolor[RGB]{216,88,106}\textcolor{black}{.631} & --- \\
uniform3 {\tiny [3,3,3,3,3,3]} & \cellcolor[RGB]{252,166,53}\textcolor{black}{18} & \cellcolor[RGB]{241,243,38}\textcolor{black}{.736} & \cellcolor[RGB]{239,248,33}\textcolor{black}{\textbf{.234}} & \cellcolor[RGB]{239,248,33}\textcolor{black}{\textbf{.301}} & \cellcolor[RGB]{140,10,164}\textcolor{white}{.907} & \cellcolor[RGB]{130,4,167}\textcolor{white}{.715} & \cellcolor[RGB]{29,6,141}\textcolor{white}{.358} \\
plus1 {\tiny [3,3,3,6,3,3]} & \cellcolor[RGB]{176,42,143}\textcolor{white}{22} & \cellcolor[RGB]{250,160,57}\textcolor{black}{.583} & --- & \cellcolor[RGB]{137,8,165}\textcolor{white}{.263} & \cellcolor[RGB]{12,7,134}\textcolor{white}{.891} & \cellcolor[RGB]{239,248,33}\textcolor{black}{\textbf{.507}} & \cellcolor[RGB]{202,70,120}\textcolor{black}{.340} \\
deep8 {\tiny [2,2,2,8,2,2]} & \cellcolor[RGB]{252,166,53}\textcolor{black}{18} & \cellcolor[RGB]{239,248,33}\textcolor{black}{\textbf{.746}} & \cellcolor[RGB]{174,39,145}\textcolor{white}{.200} & \cellcolor[RGB]{130,4,167}\textcolor{white}{.262} & \cellcolor[RGB]{24,6,139}\textcolor{white}{.892} & \cellcolor[RGB]{114,0,168}\textcolor{white}{.727} & \cellcolor[RGB]{12,7,134}\textcolor{white}{.359} \\
plus2 $^\dagger$ {\tiny [4,4,4,6,4,4]} & \cellcolor[RGB]{12,7,134}\textcolor{white}{26} & \multicolumn{1}{c}{$\times$} & \cellcolor[RGB]{126,3,167}\textcolor{white}{.192} & \cellcolor[RGB]{151,19,160}\textcolor{white}{.265} & \cellcolor[RGB]{112,0,168}\textcolor{white}{.903} & \cellcolor[RGB]{79,2,162}\textcolor{white}{.751} & \cellcolor[RGB]{230,109,90}\textcolor{black}{.335} \\
uniform4 {\tiny [4,4,4,4,4,4]} & \cellcolor[RGB]{106,0,167}\textcolor{white}{24} & \cellcolor[RGB]{244,138,71}\textcolor{black}{.537} & \cellcolor[RGB]{219,91,103}\textcolor{black}{.210} & \cellcolor[RGB]{144,14,163}\textcolor{white}{.264} & \cellcolor[RGB]{24,6,139}\textcolor{white}{.892} & \cellcolor[RGB]{12,7,134}\textcolor{white}{.789} & \cellcolor[RGB]{239,248,33}\textcolor{black}{\textbf{.321}} \\
coarse $^\dagger$ {\tiny [4,4,4,6,2,2]} & \cellcolor[RGB]{176,42,143}\textcolor{white}{22} & \multicolumn{1}{c}{$\times$} & \cellcolor[RGB]{251,162,56}\textcolor{black}{.222} & \cellcolor[RGB]{190,56,131}\textcolor{white}{.272} & \cellcolor[RGB]{240,128,77}\textcolor{black}{.929} & \cellcolor[RGB]{199,66,123}\textcolor{white}{.652} & \cellcolor[RGB]{239,248,33}\textcolor{black}{\textbf{.321}} \\
pyramid $^\dagger$ {\tiny [2,3,4,6,4,3]} & \cellcolor[RGB]{176,42,143}\textcolor{white}{22} & \multicolumn{1}{c}{$\times$} & \cellcolor[RGB]{243,135,72}\textcolor{black}{.218} & \cellcolor[RGB]{205,73,117}\textcolor{black}{.275} & \cellcolor[RGB]{239,248,33}\textcolor{black}{\textbf{.946}} & \cellcolor[RGB]{191,57,130}\textcolor{white}{.661} & \cellcolor[RGB]{250,214,36}\textcolor{black}{.324} \\
fine {\tiny [2,2,2,6,4,4]} & \cellcolor[RGB]{224,100,97}\textcolor{black}{20} & \cellcolor[RGB]{12,7,134}\textcolor{white}{$-$.070} & \cellcolor[RGB]{104,0,167}\textcolor{white}{.189} & \cellcolor[RGB]{12,7,134}\textcolor{white}{.248} & \cellcolor[RGB]{209,78,114}\textcolor{black}{.920} & \cellcolor[RGB]{169,35,149}\textcolor{white}{.683} & \cellcolor[RGB]{196,63,126}\textcolor{white}{.341} \\
\bottomrule
\end{tabular}
\end{table}

\begin{table}[H]
\centering
\small
\caption{Per-level SIGReg strength ($\lambda$ schedule) on the uniform3 backbone; the reference applies scalar $\lambda$=0.15 to L0--L3 only.}
\label{tab:app-lambd}
\setlength{\tabcolsep}{4pt}
\begin{tabular}{lcccccc}
\toprule
Variant & \rotatebox{55}{\scriptsize Chroma R² L5 ↑} & \rotatebox{55}{\scriptsize Key L5 ↑} & \rotatebox{55}{\scriptsize Root L5 ↑} & \rotatebox{55}{\scriptsize Density L5 ↑} & \rotatebox{55}{\scriptsize Cross-song min ↓} & \rotatebox{55}{\scriptsize Val loss ↓} \\
\midrule
reference {\tiny 0.15 scalar, L0–L3} & \cellcolor[RGB]{184,50,137}\textcolor{white}{.562} & \cellcolor[RGB]{253,196,39}\textcolor{black}{.208} & \cellcolor[RGB]{239,248,33}\textcolor{black}{\textbf{.267}} & \cellcolor[RGB]{158,25,156}\textcolor{white}{.790} & \cellcolor[RGB]{253,195,40}\textcolor{black}{.539} & \cellcolor[RGB]{252,170,51}\textcolor{black}{.362} \\
min {\tiny [.15$\times$4,.10,.05]} & \cellcolor[RGB]{224,100,97}\textcolor{black}{.582} & \cellcolor[RGB]{114,0,168}\textcolor{white}{.155} & \cellcolor[RGB]{121,1,168}\textcolor{white}{.226} & \cellcolor[RGB]{190,56,131}\textcolor{white}{.811} & \cellcolor[RGB]{250,156,59}\textcolor{black}{.571} & \cellcolor[RGB]{87,1,164}\textcolor{white}{.517} \\
gentle {\tiny [.15,.13,.12,.11,.10,.05]} & \cellcolor[RGB]{176,42,143}\textcolor{white}{.559} & \cellcolor[RGB]{12,7,134}\textcolor{white}{.137} & \cellcolor[RGB]{12,7,134}\textcolor{white}{.213} & \cellcolor[RGB]{242,132,75}\textcolor{black}{.860} & \cellcolor[RGB]{202,70,120}\textcolor{black}{.659} & \cellcolor[RGB]{150,18,160}\textcolor{white}{.479} \\
std {\tiny [.30,.25,.20,.15,.10,.05]} & \cellcolor[RGB]{239,248,33}\textcolor{black}{\textbf{.629}} & \cellcolor[RGB]{195,62,127}\textcolor{white}{.175} & \cellcolor[RGB]{167,33,151}\textcolor{white}{.233} & \cellcolor[RGB]{247,146,65}\textcolor{black}{.868} & \cellcolor[RGB]{239,248,33}\textcolor{black}{\textbf{.498}} & \cellcolor[RGB]{49,4,150}\textcolor{white}{.539} \\
steep {\tiny [.50,.35,.25,.15,.10,.05]} & \cellcolor[RGB]{244,234,38}\textcolor{black}{.625} & \cellcolor[RGB]{207,75,116}\textcolor{black}{.179} & \cellcolor[RGB]{223,97,99}\textcolor{black}{.245} & \cellcolor[RGB]{158,25,156}\textcolor{white}{.790} & \cellcolor[RGB]{253,196,39}\textcolor{black}{.537} & \cellcolor[RGB]{12,7,134}\textcolor{white}{.554} \\
inv1 {\tiny [.10,.13,.15,.15,.10,.05]} & \cellcolor[RGB]{214,85,109}\textcolor{black}{.576} & \cellcolor[RGB]{195,62,127}\textcolor{white}{.175} & \cellcolor[RGB]{223,97,99}\textcolor{black}{.245} & \cellcolor[RGB]{203,71,119}\textcolor{black}{.821} & \cellcolor[RGB]{253,196,39}\textcolor{black}{.537} & \cellcolor[RGB]{154,21,158}\textcolor{white}{.476} \\
inv2 {\tiny [.05,.10,.15,.20,.10,.05]} & \cellcolor[RGB]{12,7,134}\textcolor{white}{.512} & \cellcolor[RGB]{223,97,99}\textcolor{black}{.185} & \cellcolor[RGB]{250,159,58}\textcolor{black}{.255} & \cellcolor[RGB]{12,7,134}\textcolor{white}{.723} & \cellcolor[RGB]{197,64,125}\textcolor{white}{.666} & \cellcolor[RGB]{236,119,84}\textcolor{black}{.397} \\
peak $^\dagger$ {\tiny [.10,.15,.20,.20,.10,.05]} & \multicolumn{1}{c}{$\times$} & \cellcolor[RGB]{163,29,153}\textcolor{white}{.166} & \cellcolor[RGB]{149,17,161}\textcolor{white}{.230} & \cellcolor[RGB]{251,211,36}\textcolor{black}{.901} & \cellcolor[RGB]{228,106,93}\textcolor{black}{.621} & \cellcolor[RGB]{129,4,167}\textcolor{white}{.492} \\
lowflat {\tiny [.10$\times$4,.08,.05]} & \cellcolor[RGB]{253,178,47}\textcolor{black}{.609} & \cellcolor[RGB]{137,8,165}\textcolor{white}{.160} & \cellcolor[RGB]{45,4,148}\textcolor{white}{.216} & \cellcolor[RGB]{101,0,167}\textcolor{white}{.760} & \cellcolor[RGB]{193,60,128}\textcolor{white}{.671} & \cellcolor[RGB]{236,119,84}\textcolor{black}{.397} \\
lowest {\tiny [.07$\times$5,.05]} & \cellcolor[RGB]{217,89,105}\textcolor{black}{.578} & \cellcolor[RGB]{228,106,93}\textcolor{black}{.187} & \cellcolor[RGB]{202,70,120}\textcolor{black}{.240} & \cellcolor[RGB]{213,83,109}\textcolor{black}{.829} & \cellcolor[RGB]{12,7,134}\textcolor{white}{.820} & \cellcolor[RGB]{253,190,41}\textcolor{black}{.350} \\
xmep1 $^\dagger$ {\tiny cross-level MEP} & \multicolumn{1}{c}{$\times$} & \cellcolor[RGB]{239,248,33}\textcolor{black}{\textbf{.218}} & \cellcolor[RGB]{239,248,33}\textcolor{black}{\textbf{.267}} & \cellcolor[RGB]{239,248,33}\textcolor{black}{\textbf{.918}} & \cellcolor[RGB]{206,74,117}\textcolor{black}{.655} & \cellcolor[RGB]{239,248,33}\textcolor{black}{\textbf{.317}} \\
\bottomrule
\end{tabular}
\end{table}

\begin{table}[H]
\centering
\small
\caption{Chord-supervision weight $\lambda_{\mathrm{chord}}$ (heads on L0--L3 unless ``/all''; values = best across levels, one consistent probe-suite vintage; ``/all'' arms are 100-epoch screens, the rest 250 epochs).}
\label{tab:app-chordlam}
\setlength{\tabcolsep}{4pt}
\begin{tabular}{lccccc}
\toprule
$\lambda_{\mathrm{chord}}$ & \rotatebox{55}{\scriptsize Root ↑} & \rotatebox{55}{\scriptsize Key ↑} & \rotatebox{55}{\scriptsize Chroma R² ↑} & \rotatebox{55}{\scriptsize Density ↑} & \rotatebox{55}{\scriptsize Cross-song min ↓} \\
\midrule
0 {\tiny (baseline)} & \cellcolor[RGB]{12,7,134}\textcolor{white}{.229} & \cellcolor[RGB]{12,7,134}\textcolor{white}{.171} & \cellcolor[RGB]{220,94,102}\textcolor{black}{.680} & \cellcolor[RGB]{239,125,79}\textcolor{black}{.881} & \cellcolor[RGB]{12,7,134}\textcolor{white}{.836} \\
0.1 & \cellcolor[RGB]{47,4,149}\textcolor{white}{.253} & \cellcolor[RGB]{12,7,134}\textcolor{white}{.172} & \cellcolor[RGB]{12,7,134}\textcolor{white}{.460} & \cellcolor[RGB]{12,7,134}\textcolor{white}{.787} & \cellcolor[RGB]{239,125,79}\textcolor{black}{.700} \\
0.5 & \cellcolor[RGB]{248,221,36}\textcolor{black}{.602} & \cellcolor[RGB]{252,206,37}\textcolor{black}{.681} & \cellcolor[RGB]{245,231,38}\textcolor{black}{.824} & \cellcolor[RGB]{243,236,38}\textcolor{black}{.921} & \cellcolor[RGB]{71,2,159}\textcolor{white}{.813} \\
1.0 & \cellcolor[RGB]{239,248,33}\textcolor{black}{\textbf{.627}} & \cellcolor[RGB]{239,248,33}\textcolor{black}{\textbf{.737}} & \cellcolor[RGB]{239,248,33}\textcolor{black}{\textbf{.839}} & \cellcolor[RGB]{243,134,73}\textcolor{black}{.885} & \cellcolor[RGB]{212,82,110}\textcolor{black}{.728} \\
1.0 / all {\tiny (=0.167/level)} & \cellcolor[RGB]{126,3,167}\textcolor{white}{.330} & \cellcolor[RGB]{100,0,167}\textcolor{white}{.277} & \cellcolor[RGB]{242,132,75}\textcolor{black}{.726} & \cellcolor[RGB]{239,248,33}\textcolor{black}{\textbf{.925}} & \cellcolor[RGB]{227,103,95}\textcolor{black}{.714} \\
1.5 / all {\tiny (=0.25/level)} & \cellcolor[RGB]{253,196,39}\textcolor{black}{.578} & \cellcolor[RGB]{252,206,37}\textcolor{black}{.681} & \cellcolor[RGB]{245,231,38}\textcolor{black}{.824} & \cellcolor[RGB]{180,45,141}\textcolor{white}{.844} & \cellcolor[RGB]{239,248,33}\textcolor{black}{\textbf{.636}} \\
\bottomrule
\end{tabular}
\end{table}

\section{Runtime}
\label{app:runtime}
Wall-clock time for the demo pipeline (encode $\rightarrow$ sample
$\rightarrow$ render) across backends; sampling uses 10 Euler steps (20
function evaluations under classifier-free guidance). Parenthesized values are
speedups over the 2-thread CPU baseline. Yellow = fastest in row.

The table reports general-case timings, i.e.\ guidance strength $\neq 1.0$,
where each Euler step evaluates both guidance branches. At the default
strength of exactly $1.0$ the unconditional branch is skipped and sampling
times halve, matching the timings reported in the main text.

\begin{table}[H]
\centering
\small
\caption{Pipeline timing by backend.}
\label{tab:app-timing}
\setlength{\tabcolsep}{4pt}
\begin{tabular}{lcccc}
\toprule
 & CPU {\tiny 2 threads} & CPU {\tiny all cores} & MPS {\tiny (M1 Max)} & CUDA {\tiny (4090 laptop)} \\
\midrule
encode (ms) & \cellcolor[RGB]{252,167,53}\textcolor{black}{8.6 {\tiny (1$\times$)}} & \cellcolor[RGB]{251,164,54}\textcolor{black}{8.7 {\tiny (0.99$\times$)}} & \cellcolor[RGB]{12,7,134}\textcolor{white}{19.9 {\tiny (0.43$\times$)}} & \cellcolor[RGB]{239,248,33}\textcolor{black}{\textbf{5.8} {\tiny (1.49$\times$)}} \\
sample (ms) & \cellcolor[RGB]{12,7,134}\textcolor{white}{7,475 {\tiny (1$\times$)}} & \cellcolor[RGB]{129,4,167}\textcolor{white}{5,586 {\tiny (1.34$\times$)}} & \cellcolor[RGB]{253,192,41}\textcolor{black}{1,178 {\tiny (6.34$\times$)}} & \cellcolor[RGB]{239,248,33}\textcolor{black}{\textbf{194} {\tiny (38.6$\times$)}} \\
render (ms) & \cellcolor[RGB]{12,7,134}\textcolor{white}{2.7} & \cellcolor[RGB]{251,209,36}\textcolor{black}{1.7} & \cellcolor[RGB]{245,141,69}\textcolor{black}{1.9} & \cellcolor[RGB]{239,248,33}\textcolor{black}{\textbf{1.6}} \\
end-to-end (s) & \cellcolor[RGB]{12,7,134}\textcolor{white}{7.49 {\tiny (1$\times$)}} & \cellcolor[RGB]{129,4,167}\textcolor{white}{5.6 {\tiny (1.34$\times$)}} & \cellcolor[RGB]{253,190,41}\textcolor{black}{1.2 {\tiny (6.24$\times$)}} & \cellcolor[RGB]{239,248,33}\textcolor{black}{\textbf{0.2} {\tiny (37.2$\times$)}} \\
\bottomrule
\end{tabular}
\end{table}

\end{document}